%% file: cjh.tex
\documentclass[journal, twocolumn]{IEEEtran}

\input{template/package}

\title{V2X-Assisted Distributed Computing and Control Framework for Connected and Automated Vehicles under Ramp Merging Scenario}

\author{

{
   Qiong Wu,~\IEEEmembership{Senior Member,~IEEE}, 
   Jiahou Chu, 
   Pingyi Fan,~\IEEEmembership{Senior Member,~IEEE}, 
   \\
   Kezhi Wang,~\IEEEmembership{Senior Member,~IEEE},
     Nan Cheng,~\IEEEmembership{Senior Member,~IEEE}, 
   \\
      Wen Chen,~\IEEEmembership{Senior Member,~IEEE},  
   and Khaled B. Letaief,~\IEEEmembership{Fellow,~IEEE}
}

\thanks{
{Qiong Wu and Jiahou Chu are with the School of Internet of Things Engineering, Jiangnan University, Wuxi 214122, 
China (e-mail: qiongwu@jiangnan.edu.cn, 6211924035@stu.jiangnan.edu.cn).

Pingyi Fan is with the Department of Electronic Engineering, Beijing National Research Center for Information Science and Technology, 
Tsinghua University, Beijing 100084, 
China (e-mail: fpy@tsinghua.edu.cn).

Kezhi Wang is with the Department of Computer Science, Brunel University London, Middlesex UB8 3PH, 
UK (e-mail: Kezhi.Wang@brunel.ac.uk).

Nan Cheng is with the State Key Lab. of ISN and School of Telecommunications Engineering, Xidian University, Xi'an 71007l, 
China (e-mail dr.nan.cheng@ieee.org).

Wen Chen is with the Department of Electronic Engineering, Shanghai Jiao Tong University, Shanghai 200240, 
China (e-mail: wenchen@sjtu.edu.cn).

Khaled B. Letaief is with the Department of Electrical and Computer Engineering, the Hong Kong University of Science and Technology (HKUST), 
Hong Kong (email: eekhaled@ust.hk).
}
}

}

\begin{document}

\maketitle

%%%%%%%%%%%%%%%%%%%%%%%%%%%%%%%%%%%%%%%%%%%%%%%%%%%%%%%%%%%%%%%%%%%%%%%%%%%%%%%%%%%%%%%%%%%%%%%%%%%%%%%%%%%%%%%%%%%%%%
\begin{abstract}
This paper investigates distributed computing and cooperative control of connected and automated vehicles (CAVs) in ramp merging scenario under transportation cyber-physical system. 
Firstly, a centralized cooperative trajectory planning problem is formulated subject to the safely constraints and traffic performance in ramp merging scenario,
where the trajectories of all vehicles are jointly optimized.
To get rid of the reliance on a central controller and reduce computation time, 
a distributed solution to this problem implemented among CAVs through Vehicles-to-Everything (V2X) communication is proposed.
Unlike existing method, our method 
can distribute the computational task among CAVs and carry out parallel solving through V2X communication.
Then, a multi-vehicles model predictive control (MPC) problem aimed at maximizing system stability 
and minimizing control input is formulated based on the solution of the first problem
subject to strict safety constants and input limits.
Due to these complex constraints, 
this problem becomes high-dimensional, centralized, and non-convex. 
To solve it in a short time, a decomposition and convex reformulation method,
namely distributed cooperative iterative model predictive control (DCIMPC), is proposed. 
This method leverages the communication capability of CAVs to decompose the problem, 
making full use of the computational resources on vehicles to achieve fast solutions and distributed control.
The two above problems with their corresponding solving methods form the systemic framework of the V2X assisted distributed computing and control. 
Simulations have been conducted to evaluate the framework's convergence, safety, and solving speed.
Additionally, extra experiments are conducted to validate the performance of DCIMPC. 
The results show that our method can greatly improve computation speed without sacrificing system performance.
\end{abstract}

%%%%%%%%%%%%%%%%%%%%%%%%%%%%%%%%%%%%%%%%%%%%%%%%%%%%%%%%%%%%%%%%%%%%%%%%%%%%%%%%%%%%%%%%%%%%%%%%%%%%%%%%%%%%%%%%%%%%%%
\IEEEpeerreviewmaketitle

%%%%%%%%%%%%%%%%%%%%%%%%%%%%%%%%%%%%%%%%%%%%%%%%%%%%%%%%%%%%%%%%%%%%%%%%%%%%%%%%%%%%%%%%%%%%%%%%%%%%%%%%%%%%%%%%%%%%%%
\section{Introduction}

\subsection{Background}

%首先，说由于城市中车辆数目增加导致了交通压力增大和安全性问题。尤其是在匝道合流区域，由于人类驾驶员驾驶风格不同且能够接受到的信息有限，因此车辆需要通过频繁加减速以调整相对位置和速度才能完成车流合并，这不仅降低了交通性能，也会增加能量消耗，甚至会导致交通事故。但是，自动驾驶技术的发展和cyber-physical system概念的出现为解决这一问题提供了新的思路。
%在该系统下，车辆通过V2X与场景的中心控制器建通信链路，然后通过中心式计算进行协同路径规划和控制，从而实现更平滑的匝道合流控制。
%然后，说这种匝道合流区的中心式车辆控制方案存在两个缺点。第一个是这种方案通常需要建立一个计算量很大的中心式优化问题，如果直接求解不仅无法保证效率，而且不能充分利用cyber-physical system下车辆上的分布式计算资源。第二个是无法摆脱对中心控制器的依赖，这不仅增加了部署成本，也降低了通用性。因此，分布式计算和控制被当作下一代交通系统的基础技术之一。然而，这种包含多车的路径规划或控制问题通常由于车辆间存在影响而难以分布式求解。因此，研究并设计一种匝道合流场景下基于V2X的分布式计算和控制框架对加速构建先进的交通cyber-physical system具有重要意义。

\lettrine[lines=2]{W}{ITH} the number of vehicles in urban areas continues to increase, traffic congestion is becoming more severe, and safety concerns are growing increasingly critical \cite{zhu2022merging,sarvi2007microsimulation}. 
The advancement of autonomous driving presents a promising solution to these challenges. 
However, current autonomous driving are largely based on single-vehicle intelligence, 
where each vehicle only depends on its own sensors and computing units to do environmental perception and decision-making, 
lacking inter-vehicle communication and cooperation mechanisms \cite{gruyer2017perception, w1, w2, w3, w4, w5, w6, w7, w8, w9}. 
Consequently, Vehicle-to-Everything (V2X) technology, specifically designed for vehicular communication, 
has emerged as a prominent research focus in the field of communications. 
Cellular-V2X (C-V2X), a protocol stack proposed by the 3rd Generation Partnership Project (3GPP), 
enables vehicles to achieve low-latency direct communication through sidelink channels, 
facilitating the transition from ordinary vehicles to Connected and Autonomous Vehicles (CAVs) \cite{10387423, 8907851}. 
In this context, intelligent transportation systems (ITS), composed of CAVs and other edge devices, 
are rapidly evolving as a type of cyber-physical system, 
steering traditional autonomous driving towards a more networked and collaborative approach \cite{10032660, 10402048, w10, w11, w12, w13, w14,w15,w16,w17,w18,w19}.

%In contrast, Intelligent Transportation Systems (ITS) grounded in Cyber-Physical Systems (CPS) 
%utilize Vehicle-to-Everything (V2X) to facilitate communication and information sharing among vehicles, 
%leading to the development of Connected and Automated Vehicles (CAVs) \cite{yurtsever2020survey, muhammad2020deep}. 

Enabled by V2X communication, 
it becomes feasible to co-design cooperative control algorithms and distributed computing frameworks for scenarios prone to traffic congestion and accidents, 
such as ramp merging \cite{yang2023trustworthy, saad2021advancements, miao2022does}.
Ramp merging is a common traffic scenario in which vehicles from both the main road and the ramp must frequently brake and accelerate to adjust their relative positions and speeds to merge \cite{10234718}. 
However, due to the varying driving styles of human drivers and the lack of effective communication, traffic rules with suboptimal performance must often be implemented to ensure safety. 
For example, vehicles on the ramp are typically required to slow down and wait for a safe gap in the main road traffic \cite{w20,w21,w22,w23,w24,w25,w26}. 
In contrast, within a transportation CPS, CAVs can establish communication links with a central controller via V2X technology \cite{10274112}. 
This enables centralized algorithm of cooperative trajectory planning and control, resulting in smoother ramp merging.

Nevertheless, there are two significant drawbacks to this centralized vehicle control algorithms for ramp merging areas. 
Firstly, this method usually involves formulating a centralized optimization problem, which is always a computationally intensive task. 
Directly solving this problem not only fails to guarantee efficiency but also cannot fully leverage the distributed computing resources available on vehicles within the CPS \cite{9403939, 9373962, 9877926}. 
Secondly, the dependence on a central controller increases deployment costs and reduces flexibility \cite{8672604, w27,w28,w29}. 
Consequently, distributed computing and control are viewed as foundational technologies for next-generation traffic systems. 
However, these trajectory planning or control problems involving multiple vehicles are often challenging to solve in a distributed manner due to inter-vehicle interactions \cite{4522625}. 
Therefore, the research and design of a V2X-assisted distributed computing and control framework for ramp merging scenarios is of substantial importance for advancing the development of ITS \cite{8778746, 8761966}.

\subsection{Related Work and Motivation}

Currently, many works have proposed centralized computing framework for vehicle planning and control schemes based on V2X for ramp merging areas. 
In \cite{chen2023integrated}, \Athr{Chen} proposed a method for merging the traffic from the main road and the ramp into a unified flow using a mixed-integer nonlinear programming model. 
This method can determine the merging sequence while planning trajectories, thereby improving traffic performance. 
In \cite{jing2019cooperative}, \Athr{Jing} introduced a ramp merging scheme based on multiplayer cooperative game theory, considering passengers' comfort, energy consumption, and travel time as optimization objectives. 
They decomposed a multiplayer cooperative game problem into multiple two-player game problems and solved them. 
Similarly employing cooperative game theory to address multi-vehicle cooperation in ramp scenarios, \Athr{Yang} considered multi-lane situations compared to \cite{jing2019cooperative} in \cite{yang2023multi}. 
\Athr{Rios} formulated the multi-vehicle control problem in the ramp merging zone with safety constraints as an optimal control problem with minimized energy consumption and solved it using Hamiltonian analysis in \cite{rios2016automated}. 
\Athr{Kherroubi} incorporated machine learning into the ramp merging vehicle control problem in \cite{el2021novel}. 
They proposed a method using artificial neural networks to predict the trajectories of vehicles driven by human drivers, inputting the predicted trajectories into controllers trained by reinforcement learning, and then sending the controller's output to autonomous vehicles.
In \cite{hou2023cooperative}, \Athr{Hou} proposed a two-layer ramp merging control framework. 
The upper layer of the framework uses RSU to centrally calculate the merging sequence of vehicles, and then RSU sends this information to the vehicles.
In \cite{milanes2010automated}, \Athr{Milanes} proposed a two-layer control architecture for ramp merging scenarios, transforming the cooperative control problem of three vehicles into an optimal control problem with multiple safety constraints. 
They utilized the Pontryagin's Minimum Principle and Control Barrier Functions (CBFs) to solve the problem. 
In \cite{beheshtitabar2020rule}, \Athr{Beheshtitabar} studied a rule-based control algorithm, assuming dedicated lanes for convoy and non-convoy vehicles in ramp scenarios. 
Each autonomous vehicle entering the ramp must choose appropriate actions based on the traffic conditions of the main road, such as deceleration or joining a convoy, to complete the merge. 
In \cite{chen2021connected}, \Athr{Chen} proposed a virtual rotation method to formulate the ramp merging scenario as a virtual platooning problem. 
Building on this concept, the authors introduced a multi-vehicle virtual platooning model. 
Numerical simulation experiments demonstrated that their proposed control method effectively reduces gaps during merging while ensuring safety. 
Additionally, the method exhibits good adaptability to ramps of various shapes. 
In \cite{meng2023spatial}, \Athr{Meng} modeled the control of main road and ramp vehicles as two linked optimization control problems. 
To address real-time environmental changes, they solved the problem within a recursive framework.

All the above works are centralized control schemes, requiring the deployment of a base station or RSU in the scenario. 
Some work has also studied distributed vehicle control schemes for ramp merging areas. 
In these works, some are rule-based, where ramp vehicles, based on surrounding information, choose merging positions on the main road according to certain rules.
For example, in \cite{zhou2018optimal} and \cite{liu2023safety}, \Athr{Zhou} and \Athr{Liu} proposed two similar vehicle control methods, in which vehicles from the ramp are connected to the main road according to certain rules and send merging requests. 
In \cite{hwang2022autonomous}, \Athr{Hwang} first established a vehicle state control framework based on a Finite-State Machine (FSM), where vehicles switch states within the FSM based on different surrounding environments. 
When transitioning to a lane-changing state, vehicles are controlled using a controller trained through reinforcement learning. However, in this approach, lane-changing decisions for vehicles are controlled by the FSM, resulting in lower flexibility. 
Differing from the works in \cite{zhou2018optimal,liu2023safety,hwang2022autonomous}, \Athr{Xue} proposed a distributed framework based on optimization theory in \cite{xue2022platoon}.
In this framework, the upper layer is used to select a suitable merging position, and the lower layer is used to adjust the specific relative speeds and positions between vehicles. 
Furthermore, some works have discussed distributed control scheme for ramp merging areas based on machine learning.
In \cite{chen2023deep}, \Athr{Chen} considered the coexistence of autonomous vehicles and human-driven vehicles in ramp merging scenarios. 
In the study, the authors trained a vehicle controller based on Multi-Agent Reinforcement Learning (MARL), with the controller outputting discrete actions of acceleration, deceleration, and lane-changing.

In summary, centralized ramp merging frameworks are primarily based on establishing a centralized problem and solving it through a central controller, resulting in better performance compared to distributed solutions. 
However, distributed frameworks can operate independently of a central controller, offering greater adaptability. 
How to combine the advantages of centralized and distributed frameworks and avoid their drawbacks is a great challenge in the complex scenarios of ramp merging.

\begin{figure*}[htbp]
\centering 
\includegraphics[scale=0.9]{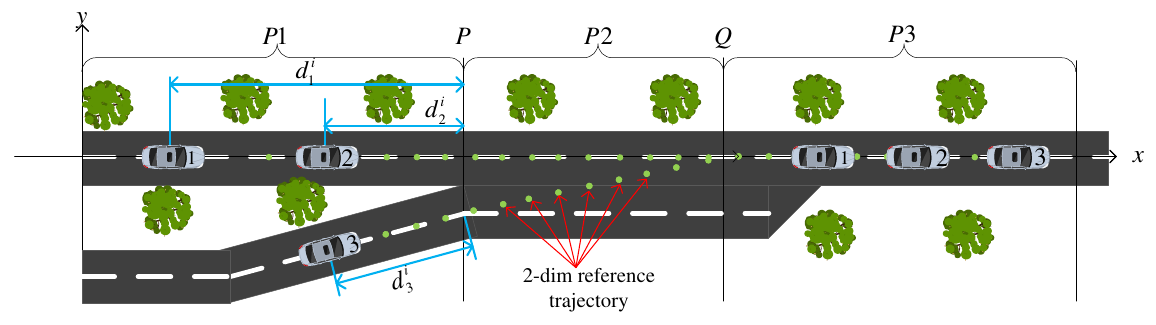}
\caption{System Model in Ramp Merging Area}
\label{system-model}
\end{figure*}

\subsection{Contributions}

To the authors' best knowledge, V2X assisted frameworks of distributed computing to solve the centralized trajectory planning and control problem in ramp merging scenarios has not been studied thus far.
In this paper, we try to find a new way to this complex problem and propose a framework where centralized optimization problems are solved by distributed computing. 
%In this paper, we propose a framework where centralized optimization problems are solved by distributed computing. 
%Simultaneously, within the optimization problem, vehicles establish collaborative relationships through safety constraints.
The main contributions of this are are summarized as follows:

\begin{itemize}
\item[1)] A trajectory planning problem and a model predictive control problem on all vehicles in the scenario have been established. 
Taking into account the characteristics of ramp merging scenarios, we first establish the longitudinal trajectory planning problem. 
This problem aims to minimize the energy consumption of all vehicles and can ensure safe distances between vehicles at a macro level. 
Based on the solution to the trajectory planning problem, we formulate a multi-vehicle model predictive control problem and incorporate safety constraints. 
This problem is used to further ensure vehicle safety at a micro level. 
These two problems together form framework proposed in this paper.

\item[2)] 
An algorithm capable of distributed computing of the trajectory planning task is proposed. 
In the trajectory planning problem, although the objective functions of each vehicle are separable, 
the presence of safety constraints prevents them from being solved in a distributed manner. 
Therefore, this paper first uses the Lagrange dual theory to derive its dual problem. 
Considering that this dual problem is a global variable consensus problem, 
a distributed solving method based on the dual ADMM algorithm is proposed, 
which can synchronously recover the solution to the original problem while solving the dual problem. 
In this method, vehicles exchange information through V2X communication and can iteratively solve the problem in a short period of time.

\item[3)] 
A method for parallel, distributed, and rapid solving of multi-vehicle MPC problem is proposed. 
In the control problem at the lower level of the framework, the presence of safety constraints turns the problem into a high-dimensional non-convex optimization problem. 
Since the control problems of vehicles need to be solved quickly within a short period of time. 
Therefore, we first propose a decomposition method to the problem based on nominal trajectories and V2X communication. 
Subsequently, we relax and convexify the safety constraints. 
Finally, each decomposed sub-problem is transformed into a low-dimensional convex quadratic programming problem, 
enabling each vehicle to solve the problem in parallel way by using its own computing resource while ensuring driving safety.
% 一种能够并行、分布式且快速求解多车模型预测控制的方法被提出。在框架下层的控制问题中，安全约束的存在使得该问题成为一个高纬度的非凸优化问题。由于车辆的控制问题需要在短时间内快速求解。因此，我们首先于标称轨迹和V2X提出了一种问题拆解方法。然后，对安全约束进行松弛和凸重构。最终，将拆解后的每个子问题转化为低维度的凸二次规划问题，使得每辆车能够在本地并行且快速求解并保证驾驶安全。

\item[4)]
The extensive simulation experiments demonstrate the effectiveness of the proposed framework, 
which is built through the joint design of control and computation, in ramp merging areas
\footnote{The code of this paper can be found at https://github.com/qiongwu86/V2X-Assisted-Distributed-Computing-and-Control-Framework-for-Connected-and-Automated-Vehicles.git}. 
Moreover, additional simulations confirm the effectiveness of DCIMPC in various scenarios. 
In the first part of the simulations, we investigate several aspects of the proposed framework, 
such as its effectiveness, convergence, and convergence speed. 
In the second part, we conduct additional validation experiments for DCIMPC in intersection and T-junction scenarios, 
demonstrating that this distributed computing method can achieve efficient and safe distributed control.

\end{itemize}

The rest of this paper is organized as follows. 
% 系统模型、路径规划问题和控制问题的建立在第二章中介绍。路径规划问题的分布式求解方法在第三章中介绍。然后，在第四章中介绍用于求解多车控制问题的DCIMPC框架。
System models, the establishment of trajectory planning problems and control problems is introduced in \SecRef{sec-2}. 
The distributed solving method for trajectory planning problems is presented in \SecRef{sec-3}. 
Subsequently, the DCIMPC framework for solving multi-vehicle control problems is discussed in \SecRef{sec-4}.
In Section \SecRef{sec-5}, simulation results are provided and discussed, and finally, the paper is concluded in \SecRef{sec-6}.

\emph{Notations}:
Throughout this paper, for any vector $X$ and diagonal matrix $Q$, $\left\| X \right\|^2$ and $\left\| X \right\|^2_Q$ denoted as $X^TX$ and $X^TQX$, respectively.
$\left\langle X, Y \right\langle$ denotes the inner product of vector $X$ and $Y$.
$R^{a}$ and $R^{a \times b}$ denotes the set of $a$ read-valued vectors and $a \times b$ real-valued matrix, respectively.
$\frac{\partial F}{\partial X}$ denotes the the partial derivative of $X$ to $F$.
$E^k$ and $0^k$ denotes the identity matrix and zero matrix of $k$, and $1^{a \times b}$ denotes the matrix of size $a \times b$ with all elements equal to 1.
$\otimes$ denotes Kronecker product.
For any matrix $M$, $M[a:]$, $M[:b]$, and $M[a:b]$ represent the submatrices of $M$ corresponding to the elements from the $a$ row to the last row, 
from the first row to the $b$ row, and from the $a$ row to the $b$ row, respectively.
For any vector $x$, $diag\{x\}$ denotes the a diagonal matrix with $x$ as its diagonal elements.

%%%%%%%%%%%%%%%%%%%%%%%%%%%%%%%%%%%%%%%%%%%%%%%%%%%%%%%%%%%%%%%%%%%%%%%%%%%%%%%%%%%%%%%%%%%%%%%%%%%%%%%%%%%%%%%%%%%%%%
\section{System Model}\label{sec-2}

We will introduce the system model of this paper.
In the first subsection, 
we will describe the multi-vehicle ramp merging problem 
and the overview of the distributed computing and control framework proposed in this paper, 
which solves the problem in two steps.
Then, the centralized optimization problems corresponding to these two steps 
will be formulated in \SecRef{section-2-1} and \SecRef{sec-2-2}, respectively. 
The detailed solution methods for these problems will be provided in later sections.

\subsection{Ramp Merging Problem}

\begin{figure}
\centering
\includegraphics[scale=1.0]{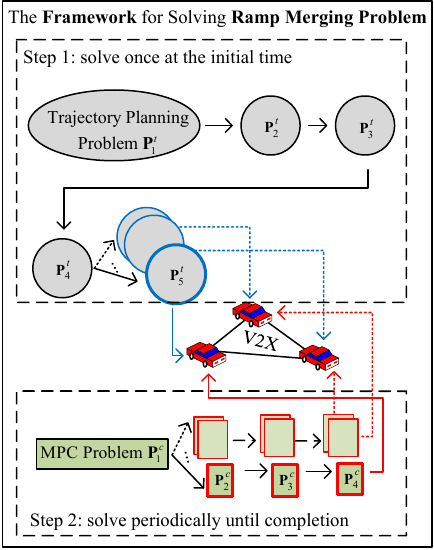}
\caption{The framework proposed in this paper.}
\label{frameworks}
\end{figure}

Consider a ramp merging scenario that contains a number of CAVs  
and all of them can communicate with one another by V2X as shown in \FigRef{system-model}.
The vehicles, comming from both the main road and the ramp, 
will communicate to each other and utilize its limited computing resource to perform distributed 
control to complete the merging process and pass through the ramp merging area quickly while ensuring safety.

% \wNew{
% Then, all vehicles in the scenario will perform distributed \uline{control and computations} using the algorithm presented in this work, 
% assisted by \uline{V2X communication}, to achieve efficient and safe ramp merging.
% }
Specifically, the entire ramp merging scenario is divided into three regions: $P1$, $P2$, and $P3$. 
Vehicles are spawned in region $P1$, then pass through region $P2$, 
and finally enter region $P3$ to complete the ramp merging. 
To enhance traffic performance for such a multi-vehicle control problem, 
it is essential to first maintain overall traffic flow stability at the macroscopic level to achieve smooth merging. 
Secondly, at the microscopic level, it is necessary to consider the kinematic model and ensure driving safety. 
Therefore, we divide the entire ramp merging problem into two steps, as illustrated in \FigRef{frameworks}.

In the first step, we formulate a centralized multi-vehicle trajectories planning problem $\mathbf{P}_1^t$ at the macroscopic level in \SecRef{section-2-1}, 
which is established and solved once at the initial moment of the merging process. 
This problem focuses on optimizing traffic flow stability and determines the merging order of vehicles as well as their terminal states. 
To ensure computational efficiency, 
this problem only solves for one-dimensional trajectories and includes relatively relaxed longitudinal safety constraints. 
It is then transformed into two-dimensional reference trajectories based on the road geometry.

In the second step, we formulate a centralized multi-vehicle model predictive control problem $\mathbf{P}_1^c$ at the microscopic level in \SecRef{sec-2-2}, 
based on the reference trajectory generated in the first step. 
This problem needs to be solved periodically throughout the entire ramp merging process. 
It incorporates a complex vehicle kinematic model and strict safety constraints to achieve low-level vehicle control.

To enable distributed computing and control, these two problems will be solved with the methods proposed later in this paper, 
leveraging V2X communication.

% A trajectory planning problem that includes all vehicles based on First In First Out (FIFO) criterion will be established, 
% which is detailed in \SecRef{section-2-1}.
% To solve this complex problem, we will do it in two step.
% In the first step, considering that this problem needs to be solved in a short period of time, 
% the safety constrain in it is relatively relaxed.
% To further reduce computation time,
% a manner for solving it through distributed computing on vehicles with the assistance of V2X communication is proposed. 
% Then, in the second step, the planned trajectory will serve as parameters for formulating a high-dimensional MPC problem,
% which is used for vehicle control and formulated in \SecRef{sec-2-2}.
% For safety, it includes strict safety constraints and needs to be solved as quickly as possible to keep higher control frequency. 
% We propose a convex reformulation method and a decomposition approach for this problem 
% to achieve distributed control and computation reducing.
% The detail parts will be given in \SecRef{sec-3} and \SecRef{sec-4}, respectively.

\subsection{Trajectory Planning Problem}\label{section-2-1}

\begin{figure}[htbp]
\centering 
\includegraphics[scale=0.8]{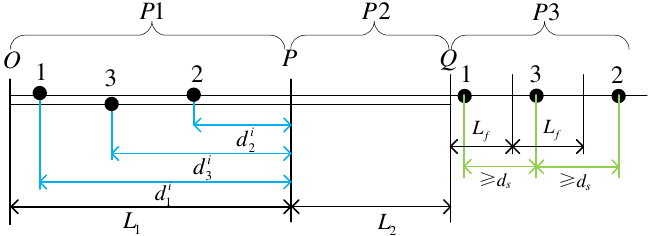}
\caption{1-dim abstract model of ramp merging area}
\label{one-dim system-model}
\end{figure}

The one-dimensional longitudinal vehicle movement model used in this section is a first-order lag model \cite{katriniok2017distributed}, which can be expressed as ${{\dot{X}}^{i}}(t)=A{{X}^{i}}(t)+B{{U}^{i}}(t)$, and
\begin{equation}\begin{aligned}
A=\begin{bmatrix}
   0 & 1 & 0  \\
   0 & 0 & 1  \\
   0 & 0 & -1/{{T}_{l}}  \\
\end{bmatrix}, 
B=\begin{bmatrix}
  0 \\ 
  0 \\ 
  1/{{T}_{l}} 
\end{bmatrix}
\end{aligned}\label{equ-1},\end{equation}
where ${{T}_{l}}$ represents the time constant of the lag, the superscript $i$ denotes the vehicle index, and there are a total of $N$ vehicles. 
The state variables ${{X}^{i}}$ and control variable ${{U}^{i}}$ are defined as
\begin{equation}\begin{aligned}
{{X}^{i}}(t)=\begin{bmatrix}
  {{s}^{i}}(t) \\ 
  {{v}^{i}}(t) \\ 
  {{a}^{i}}(t) \\ 
\end{bmatrix}, 
{{U}^{i}}(t)=\begin{bmatrix}
  a_{ref}^{i}(t) 
\end{bmatrix},
\end{aligned}\label{equ-2}\end{equation}
where ${{s}^{i}}(t)$, ${{v}^{i}}(t)$, and ${{a}^{i}}(t)$ represent the longitudinal position, velocity, and acceleration of the vehicle $i$, respectively, and $a_{ref}^{i}(t)$ is the input reference acceleration. 
Assuming a sampling time interval of ${{T}_{s}}$ is adopted, the kinematic model is turn into a discrete mode using zero-order hold (ZOH) technique as
\begin{equation}\begin{aligned}
X_{k+1}^{i}=\bar{A}X_{k}^{i}+\bar{B}U_{k}^{i}
\end{aligned}\label{equ-3},\end{equation}
where $\bar{A}$ and $\bar{B}$ are the discretized system matrix and input matrix, respectively.
In addition, $k\in \left[ 0,T_{P}^{1}-1 \right]$ is the discrete time index, and $T_{P}^{1}$ is the time steps for trajectory planning. 
% Since \EquRef{equ-1} represents a linear time-invariant system, $\bar{A}$ and $\bar{B}$ are constant matrices. 
The vehicles are projected onto a one-dimensional longitudinal coordinate axis, as shown in \FigRef{one-dim system-model}. 
Assuming the initial state of vehicle $i$ is $X_{0}^{i}$, and the order of entry into area $P3$ is denoted by ${{\mathcal{S}}_{i}}$, an integer ranging from 1 to $N$ indicating the priority sequence. 
For safety and stability considerations, all vehicles should be at suitable and safe longitudinal positions when entering area $P3$, which is constrained by the longitudinal position at the termination time $T_{P}^{1}$ as
\begin{equation}\begin{aligned}
({{L}_{1}}+{{L}_{2}})+({{\mathcal{S}}_{i}}-1) {{L}_{f}}\le s_{{{T}_{p}}}^{i}\le ({{L}_{1}}+{{L}_{2}})+{{\mathcal{S}}_{i}} {{L}_{f}},
\end{aligned}\label{equ-4}\end{equation}
where ${L}_{1}$ and ${L}_{2}$ represent the longitudinal lengths of areas $P1$ and $P2$, 
and ${L}_{f}$ denotes the range of longitudinal positions for each vehicle at time step $T_{P}^{1}$, 
as depicted in \FigRef{one-dim system-model}.

The merging occurs in area $P2$, where vehicle $i$ should always maintain a safe distance from the vehicles in front and behind it in the lane. 
The relationships between the front and back vehicles of vehicle $i$ change before and after merging. 
For example, $V1$, $V2$ and $V3$ represent three vehicles, 
which are spawned on region $P1$ and we label them with number on its icon in \FigRef{system-model}.
Before the merging of $V3$, $V1$ needs to maintain a distance from $V2$, but after merging, the sequence will become $V1 \to V3 \to V2$, 
which is shown in \FigRef{one-dim system-model}. 
To establish these time-varying constraints, the merging process of vehicle $i$ is simplified to a moment $t_{M}^{i}$, and the merging process is constrained within the range of $P2$ by establishing constraints on the average speed within $k\in \left[ 0, t_{M}^{i} \right]$ as 
\begin{equation}
\frac{{L_{1} - s_{0}^{i}}}{{T_{s}}} \le \sum_{k=1}^{t_{M}^{i}}{v_{k}^{i}} \le \frac{{L_{1} + L_{2} - s_{0}^{i}}}{{T_{s}}}.
\label{equ-5}\end{equation}
In \EquRef{equ-5}, the left hand side of the inequality represents the minimum average speed of vehicle $i$ at $k = t_{M}^{i}$ when reaching the right side of point $P$, 
and the right hand side represents the maximum average speed without passing point $Q$. 
Considering that the speeds of each vehicle should be approximately equal after the planning to ensure the stability of the platoo, it is assumed that the time intervals $t_{M}^{i}$ of each vehicle are the same, calculated as
\begin{equation}
t_{M}^{i} = T_{p} - c_{1} (\mathcal{S}_{i} - 1) - c_{0},
\label{equ-6}\end{equation}
where $c_{0}$ is the number of time steps from the last vehicle merging to passing point $Q$, and $c_{1}$ is a constant interval for each vehicle $t_{M}^{i}$.

Next, safety constraints between vehicles are established based on $t_{M}^{i}$. 
When $k \le t_{M}^{i}$, 
vehicle $i$ should maintain a safe distance $d_{S}^{1}$ from its front vehicle, denoted by $I_{i}^{P}$,
in the same lane.
However, when $k > t_{M}^{i}$, 
vehicle $i$ should maintain a safe distance from the front vehicle in the merging order, denoted by $I_{i}^{Q}$. 
Therefore, the safety constraints can be expressed as
\begin{equation}\begin{aligned}
\begin{cases}
d_{S}^{1} \le s_{k}^{I_{i}^{P}} - s_{k}^{i} & k \le t_{M}^{i} \\
d_{S}^{1} \le s_{k}^{I_{i}^{Q}} - s_{k}^{i} & k > t_{M}^{i}
\end{cases}
\end{aligned}\label{equ-7}.\end{equation}

Finally, vehicles should minimize acceleration and deceleration and constrain acceleration within a certain range. 
The trajectory planning problem $\mathbf{P}_1^t$ with safe constraint of all vehicle in ramp merging scenario is formulated as
\begin{equation}
\mathbf{P}_1^t: \left.
\begin{array}{c}
\underset{U_{k}^{i}, \text{\space} i \in \mathcal{N}, \text{\space} k \in [0, T_{P}^{1}-1]}{\min} \sum_{i=1}^{N} \sum_{k=0}^{T_{P}^{1}-1} (U_{k}^{i})^{2} \\
\text{s.t.} \quad a_{\min} \le U_{i}^{k} \le a_{\max},\\
% \text{\EquRef{equ-5}}
\text{\eqref{equ-3}, \eqref{equ-4}, \eqref{equ-5}, \eqref{equ-7}},\\
\forall i \in [1,N],\ k \in [0,T_{P}^{1}-1],
\end{array}
\right.
\label{equ-8}\end{equation}
where $a_{\min}$ and $a_{\max}$ represent the lower and upper limits of the control quantity. 
% Since the $U^i_k$ is one-dimensional vector, $a_{\min} \in R$ and $a_{\max} \in R$.

\begin{remark}
Although the objective in \EquRef{equ-8} can be decomposed for each vehicle, 
the safety constraints in \EquRef{equ-7} couple the subproblems corresponding to each vehicle, 
making it impossible to solve them in a distributed manner directly. 
To get rid of dependence on centralized controller and
fully utilize vehicular computing resource, 
a V2X assisted distributed solving scheme is proposed based on Lagrange theory and ADMM algorithm in the \SecRef{sec-3}.
\end{remark}

The solution of problem in \EquRef{equ-8} represents a one-dimensional longitudinal trajectory. 
Considering the particularity of the merging area, we adopts a simple method to generate a two-dimensional reference trajectory. 
As shown in the green origin in \FigRef{system-model}, parametric equations with respect to the longitudinal position $s^i_k$ are established along the lane center of the main road and the ramp from $P1$ to $P3$. 
They are denoted as
\begin{equation}\begin{aligned}
{{P}_{main}}(s)=\left[ \begin{matrix}
   x=f_{x}^{1}(s)  \\
   y=f_{y}^{1}(s)  \\
\end{matrix} \right],{{P}_{merge}}(s)=\left[ \begin{matrix}
   x=f_{x}^{1}(s)  \\
   y=f_{y}^{1}(s)  \\
\end{matrix} \right].
\end{aligned}\label{equ-9}\end{equation}

After solving \EquRef{equ-8} to obtain ${U^{{i*}}}$, vehicle $i$ first generates $X_{k}^{i}$ for $k\in [0,T_{P}^{1}]$ based on \EquRef{equ-3}, then take out their first dimension to form the sequence ${[\begin{matrix}
   s_{0}^{i} & s_{1}^{i} & \ldots  & s_{T_{P}^{1}}^{i}  \\
\end{matrix}]}^{T}$, and input it into ${{P}_{main}}(s)$ or ${{P}_{merge}}(s)$ to obtain a two-dimensional reference trajectory with respect to the sampling points. 
As shown in \FigRef{system-model}, the green sampling points distributed along the road center marked by red arrows represent the two-dimensional reference trajectory. 
There is a time interval of ${T}_{S}$ seconds between two adjacent points on the trajectory. 
It is important to note that the two-dimensional reference trajectory at this point can only ensure the longitudinal safety of vehicles in areas $P1$ and $P3$, but not in the merging area, i.e., area $P2$, which will be addressed in the lower-level vehicle control.

\subsection{Multi-Vehicle MPC Considering Safety Constraints}\label{sec-2-2}

\begin{figure}[htbp]
\centering 
\includegraphics[scale=0.8]{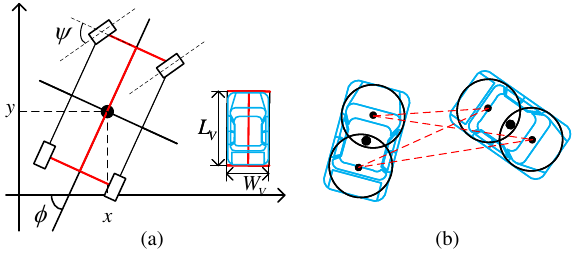}
\caption{kinematic model (a) and circular approximation (b)}
\label{kinematic-model}
\end{figure}

This subsection establishes the vehicle control problem based on the solution of 
trajectory planning problem formulated last subsection, 
further ensuring driving safety. 
The kinematic model of the vehicle can be represented as $\dot{X} = f_{S}(X, U)$ \cite{bae2020cooperation}. 
The state variable $X$, control variable $U$, and system function $f_{S}$ are defined as
\begin{equation}\begin{aligned}
X = \begin{bmatrix}
   x \\
   y \\
   \phi \\
   v
\end{bmatrix}, 
U = \begin{bmatrix}
   a \\
   \psi
\end{bmatrix}, 
f_{S}(X, U) = \begin{bmatrix}
   v\cos(\phi + \beta) \\
   v\cos(\phi + \beta) \\
   v\frac{\sin(\beta)}{0.5 \times L_{V}} \\
   a
\end{bmatrix},
\end{aligned}\label{equ-10}\end{equation}
and $\beta$ is defined as
\begin{equation}
\beta = \arctan(0.5 \times \tan(\psi)),
\label{equ-11}\end{equation}
where $x$ and $y$ represent the lateral and longitudinal coordinates of the vehicle's center point, $\phi$ is the heading angle, $v$ is the speed scalar, $a$ is the vehicle's acceleration, and $\psi$ is the front wheel steering angle as shown in \FigRef{kinematic-model} of the vehicle kinematic model. 
This subsection adopts a discretized MPC with the same sampling interval as the trajectory planning, i.e., $T_{S}$. 
Let $k$ be the discrete sampling point index, then we have $X^{i}(k \times T_{S}) = X_{k}^{i}$. 
By applying forward differencing to kinematic model, the discretization yields
\begin{equation}
\dot{X}_{k}^{i} = f_{S}(X_{k}^{i}, U_{k}^{i}) \approx \frac{X_{k+1}^{i} - X_{k}^{i}}{T_{S}}
\label{equ-12},\end{equation}
and the discrete system function is given by:
\begin{equation}
X_{k+1}^{i} = f_{d}(X_{k}^{i}, U_{k}^{i}) = T_{S} f_{S}(X_{k}^{i}, U_{k}^{i}) + X_{k}^{i}.
\label{equ-13}\end{equation}

In MPC, the controller solves an optimization problem at time step $k$ to compute the next $T_P^2$ control inputs, but only applies the first input to the system. 
When reaching time step $k+1$, the prediction horizon is shifted by one sampling interval and the process is repeated. 
The symbols $U_{(k+l,k)}^i$, $l\in [0,T_{P}^{2}-1]$, represent the subsequent $T_P^2$ control variables to be solved at time $k$. 
Similarly, $X_{(k+l,k)}^i$, $l\in [1,T_P^2]$, represent the subsequent $T_P^2$ states predicted based on $U_{(k+l,k)}^i$. 
Furthermore, for consistency in notation, the state of vehicle $i$ at time $k$ is denoted as $X_{(k,k)}^{i}$, which is a known and determined vector at time $k$. 

Based on the characteristics of MPC, safety constraints are introduced into the predicted trajectories of vehicles. 
This means that for $\forall l\in [1,T_P^2]$ and $\left( i,j \right)\in \{(a,b):a,b\in \mathcal{N}\bigcap a\ne b\}$, there should be a certain distance constraint satisfied between $X_{(k+l,k)}^i$ and $X_{(k+l,k)}^j$. 
Therefore, the MPC problem $\mathbf{P}_1^c$ with safety constraints at time $k$ can be represented as,
\begin{subequations}
\begin{align}
\;\;\; \underset{\substack{U_{\left( k+l,k \right)}^{i}, i \in \mathcal{N} \\ l \in \left[ 0, T_P^2-1 \right]}}{\min} 
& \sum_{i \in \mathcal{N}} \Bigg [ \sum_{l=1}^{T_P^2 - 1} \left\| M_f \left( X_{(k+l+1,k)}^i - X_{(k+l,k)}^i \right) \right\|^2 \nonumber
\\ 
& \!\!\!\!\!\!\!\!\!\!\!\!\!\!\!\!\!\! +\sum_{l=1}^{T_P^2} \left\| X^i_{(k+l,k)} - \tilde{X}^i_{(k+l,k)} \right\|^2_{Q_X} 
\!\!\!\!\!
+ \left\| U^i_{(k+l-1,k)} \right\|^2_{Q_U} \Bigg ] \nonumber 
\\
\mathbf{P}_1^c: \;\;\;\;\;\;\;
\text{s.t.} \;
& \underline{U} \le U_{\left( k+l,k \right)}^{i}\le \overline{U}, \label{equ-14a} \\
& X_{(k+l+1,k)}^{i} = f_d \left( X_{(k+l,k)}^{i},U_{(k+l,k)}^{i} \right), \label{equ-14b}\\
& D_s \le D \left( X_{(k+l+1,k)}^{i},X_{(k+l+1,k)}^{j} \right), \label{equ-14c} \\
& \forall i \in \mathcal{N}, l \in \left[ 0, T_P^2-1 \right], j \in \mathcal{N}_i. \nonumber
\end{align}
\label{equ-14}\end{subequations}

In the objective in \EquRef{equ-14}, 
the first item represents the rate of state changing, 
such as speed and heading angle changes. 
This item can stabilize the system, where the weight of it is controlled by the matrix ${M_{f}}$. 
The second item is to measure the difference between the vehicle and the reference trajectory within the subsequent $T_{P}^{2}$ prediction steps, which is controlled by the symmetric matrix ${Q_{X}}$.
The third item is a cost function related to control variables, which is controlled by the symmetric matrix ${Q_{U}}$.

As for the constraint conditions in \EquRef{equ-14}, \EquRef{equ-14a} pertains to the constraints on control variables, where $\overline{U}$ and $\underline{U}$ represent the upper and lower limits of control variables, respectively. 
\EquRef{equ-14b} involves discretized dynamic constraints. 
\EquRef{equ-14c} is safety constraints, where ${\mathcal{N}_i}$ denotes neighboring vehicles of vehicle $i$, defined as other vehicles in $\mathcal{N}$ excluding $i$. 
Safety constraints are established based on the trajectories predicted by MPC for each vehicle, and $D(\cdot,\cdot)$ in \EquRef{equ-14c} is a function used to measure the distance between trajectories\cite{bae2020cooperation}, 
as shown on the right side of \FigRef{kinematic-model}.
Specifically, the formula for function $D(\cdot,\cdot)$ is
% When the vehicles are elongated, the number of circles can be expanded. 
% defined as $R^4 \times R^4 \rightarrow R^4$, which is
% We approximate the vehicles as two circles, defined as
\begin{equation}\begin{aligned}
&D(X_{(k+l,k)}^{i},X_{(k+l,k)}^{j})\\
=&
\begin{bmatrix}
   \dot{D}(X_{(k+l,k)}^i,X_{(k+l,k)}^j,1,1)  \\
   \dot{D}(X_{(k+l,k)}^i,X_{(k+l,k)}^j,1,-1)  \\
   \dot{D}(X_{(k+l,k)}^i,X_{(k+l,k)}^j,-1,1)  \\
   \dot{D}(X_{(k+l,k)}^i,X_{(k+l,k)}^j,-1,-1)
\end{bmatrix},
\end{aligned}\label{equ-15}\end{equation}
where $\dot{D}(\cdot,\cdot)$ is defined as
\begin{equation}\begin{aligned}
& \dot{D}(X_{(k+l,k)}^{i},X_{(k+l,k)}^{j},p,q)\\
:= & \begin{bmatrix}
x_i - x_j + d_H^K (p\times \cos{\phi_i} + q\times \cos{\phi_j}) \\ 
y_i - y_j + d_H^K (p\times \sin{\phi_i} + q\times \sin{\phi_j})
\end{bmatrix},
\end{aligned}\label{equ-16}\end{equation}
where $d_{H}^{K}=0.5\times ({L_{V}}-{W_{V}})$.
Each term in \EquRef{equ-15} is used to calculate the center distance between two circles, totaling four pairs. 

\begin{remark}
In \EquRef{equ-14}, we achieve multi-vehicle cooperative control through safety constraints, 
but this also leads to the vehicle control problem becoming a high-dimensional non-convex nonlinear optimization problem.
If it is solved directly by the central controller without any preprocessing, it would result in a computationally intensive task.
This not only relies on the central controller but also makes it difficult to guarantee the control frequency. 
To solve it quickly and fully utilize the computing resources and communication ability of the vehicles in transportation cyber-physical system,
we propose a decomposition and a convex reformulation method in \SecRef{sec-4} to this problem, 
enabling it to be solved rapidly in parallel and distributedly on each vehicle.
\end{remark}

%%%%%%%%%%%%%%%%%%%%%%%%%%%%%%%%%%%%%%%%%%%%%%%%%%%%%%%%%%%%%%%%%%%%%%%%%%%%%%%%%%%%%%%%%%%%%%%%%%%%%%%%%%%%%%%%%%%%%%
\section{Distributed trajectory planning}\label{sec-3}

\begin{figure}[htbp]
\centering 
\includegraphics[scale=0.65]{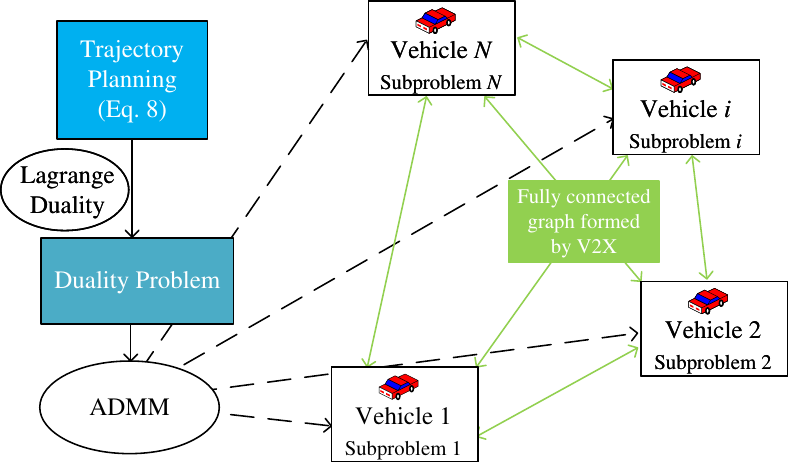}
\caption{The solving procedure of \EquRef{equ-8}}
\label{admm}
\end{figure}

The processing flow for the trajectory planning problem is illustrated in \FigRef{admm}. 
Since the problem satisfies the Slater condition, that is, the original problem and the Lagrangian dual problem satisfy strong duality with a duality gap of zero\cite{boyd2004convex}. 
Therefore, in this section, we first derive the dual problem of \EquRef{equ-14} based on the Lagrangian duality theory, 
and then uses a variant of the ADMM algorithm proposed in \cite{mateos2010distributed} to decompose the dual problem into sub-problems corresponding to each vehicle for distributed iterative solution. 
Then, vehicles will communicate through V2X and use local computing resources for distributed solving.
At the same time, by using the conclusions from \cite{banjac2019decentralized}, 
the solution to the original problem is obtained while solving the dual sub-problems.

\subsection{The dual problem of the trajectory planning problem}\label{sec-3-1}

For clarity of expression, let the initial time for trajectory planning be 0, denote the initial state vector of vehicle $i$ as $X_{0}^{i}$. 
Then, concatenate the state variables for $t\in [1,T_{P}^{1}]$ and the control variables for $t\in [0,T_{P}^{1}-1]$ together as
\begin{equation}
{{X}^{i}}=\left[ \begin{matrix}
   X_{1}^{i}  \\
   X_{2}^{i}  \\
   \vdots   \\
   X_{T_{P}^{1}}^{i}  \\
\end{matrix} \right]\in {{R}^{3T_{P}^{1}}},
{{U}^{i}}=\left[ \begin{matrix}
   U_{0}^{i}  \\
   U_{1}^{i}  \\
   \vdots   \\
   U_{T_{P}^{1}-1}^{i}  \\
\end{matrix} \right]\in {{R}^{T_{P}^{1}}} . \label{equ-17}
\end{equation}
Thus, the trajectory planning problem \EquRef{equ-8} can be rewritten as
\begin{equation}
\mathbf{P}_2^t: \left.
   \begin{aligned}
   \underset{{{U}^{i}},{{X}^{i}},i\in \mathcal{N}}{\mathop{\min }}\,\sum\limits_{i\in \mathcal{N}}{{{F}^{i}}({{U}^{i}},{{X}^{i}})}  \\
   \text{s.t.} \sum\limits_{i\in \mathcal{N}}{({{G}^{i}}{{X}^{i}}-{{H}^{i}})\in \mathcal{K}},
   \end{aligned}
\right.
\label{equ-18}
\end{equation}
where ${F}_{i}({U}_{i},{X}_{i})$ is the local objective function for vehicle $i$, defined as
\begin{equation}
{{F}^{i}}({{U}^{i}},{{X}^{i}})={{\left\| {{U}^{i}} \right\|}^{2}}+{{\mathcal{I}}_{{{C}^{i}}}}({{U}^{i}},{{X}^{i}}).
\label{equ-19}
\end{equation}
${{\mathcal{I}}_{{{C}^{i}}}}({{U}^{i}},{{X}^{i}})$ in \EquRef{equ-19} is an indicator function, which is defined as
\begin{equation}
{{\mathcal{I}}_{{{C}^{i}}}}({{U}^{i}},{{X}^{i}})=
\left\{ 
   \begin{aligned}
   0, & \quad ({{U}^{i}},{{X}^{i}})\in {{C}^{i}},  \\
   \infty,  & \quad \text{else}.
   \end{aligned}
\right.
\label{equ-20}
\end{equation}
By this function, each vehicle's own constraint conditions are incorporated into its objective function ${F}^{i}$. 
Therefore, in \EquRef{equ-19}, the set $C^i$ is the feasible domain of the control variables $U^i$ and state variables $X^i$ for the other constraints in \EquRef{equ-8}, except for the safety constraints, i.e.,
\begin{equation}
{{C}^{i}}:=\left\{ ({{U}^{i}},{{X}^{i}}):
\begin{aligned}
   & \forall k\in [0,T_{P}^{1}-1],\underline{U}\le U_{k}^{i}\le \overline{U}  \\
   & \eqref{equ-3} \cap \eqref{equ-4} \cap \eqref{equ-5}
\end{aligned} \right\}. \label{21}
\end{equation}
The matrix ${G}_{i}\in {{R}^{N{{T}_{P}}\times 3{{T}_{P}}}}$, ${{H}_{i}}\in {{R}^{N{{T}_{P}}}}$, and the convex cone $\mathcal{K}$ are used to express the safety constraints constructed in \EquRef{equ-7}. 
Specifically, some elements in $G^i$ satisfy 
\begin{subequations}
\begin{align}
& {{G}^{i}}[i T_{P}^{1}:(i+1) T_{P}^{1}]=
\begin{bmatrix} -1 & 0 & 0 \end{bmatrix}
\otimes {{E}^{T_{P}^{1}}} ,\label{equ-22a} \\ 
& {{G}^{i}}[J_{i}^{P} T_{P}^{1}:J_{i}^{P} T_{P}^{1}+t_{J_{i}^{P}}^{M}]=
\begin{bmatrix} 1 & 0 & 0  \end{bmatrix}
\otimes {{E}^{T_{P}^{1}}}[:t_{J_{i}^{P}}^{M}] , \label{equ-22b}\\ 
& \begin{aligned}
&{{G}^{i}}[J_{i}^{Q} T_{P}^{1}+t_{J_{i}^{Q}}^{M}:(J_{i}^{Q}+1) T_{P}^{1}] \\ 
&=\begin{bmatrix} 1 & 0 & 0 \end{bmatrix}
\otimes {{E}^{T_{P}^{1}}}[t_{J_{i}^{Q}}^{M}:] ,
\end{aligned}\label{equ-22c}
\end{align}
\end{subequations}
with all others being 0 elements.
\EquRef{equ-22a} extracts the first dimension (longitudinal coordinate) of ${X}_{i}$, which is used to express the safety constraints when vehicle $i$ is the following vehicle. 
\EquRef{equ-22b} and \EquRef{equ-22c} are used to express the safety constraints for vehicle $i$ when it is the leading vehicle. 
Specifically, as mentioned in \SecRef{sec-2-2}, vehicle $i$ could either be the leading vehicle before a car passes point $P$ or the leading vehicle after another car passes point $Q$. 
To distinguish from the leading vehicle of vehicle $i$, we use $J_{i}^{P}$ and $J_{i}^{Q}$ to represent the numbers of these two vehicles. 
Then, in equation \EquRef{equ-18}, ${H}^{i}$ satisfies
\begin{equation}
   H^i = \begin{bmatrix}
      0^{1 \times (i-1)T_P^1}& d_s& \cdots& d_s& 0^{1 \times (N-i)T_P^1}
   \end{bmatrix}^T.
\label{equ-23}\end{equation}
The convex cone $\mathcal{K}$ is a positive quadrant cone on the space ${{R}^{NT_{P}^{1}}}$, i.e., it is a subset formed by all components in the set ${{R}^{NT_{P}^{1}}}$ being greater than 0, denoted as ${{R}^{NT_{P}^{1}+}}$. 
Considering that \EquRef{equ-3} is a linear time-invariant system, the dynamic constraints can be expressed in terms of ${X}_{i}$ and ${U}_{i}$ to simplify the expression.
Due to the recursive nature of the system dynamics,
\begin{equation}
\begin{aligned}
X_{1}^{i}&=\bar{A}X_{0}^{i}+\bar{B}U_{0}^{i}  , \\
X_{2}^{i}&=\bar{A}X_{1}^{i}+\bar{B}U_{1}^{i}={{{\bar{A}}}^{2}}X_{0}^{i}+\bar{A}\bar{B}U_{0}^{i}+\bar{B}U_{1}^{i} , \\ 
% & X_{3}^{i}=\bar{A}X_{2}^{i}+\bar{B}U_{2}^{i}={{{\bar{A}}}^{3}}X_{0}^{i}+{{{\bar{A}}}^{2}}\bar{B}U_{0}^{i}+\bar{A}\bar{B}U_{1}^{i}+\bar{B}U_{2}^{i} \\ 
&\vdots   \\
X_{{{T}_{P}}}^{i}&={{{\bar{A}}}^{T_{P}^{1}}}X_{0}^{i}+\sum\limits_{k=0}^{T_{P}^{1}-1}{{{{\bar{A}}}^{(T_{P}^{1}-1-k)}}\bar{B}U_{k}^{i}}.
\end{aligned}
\label{equ-24}\end{equation}
Therefore, it follows that
\begin{equation}{{X}^{i}}=\Gamma X_{0}^{i}+\Lambda {{U}^{i}} , \label{equ-25} \end{equation}
where
\begin{equation}
\Gamma =
\begin{bmatrix}
{\bar{A}} \\ {{{\bar{A}}}^{2}} \\ \vdots \\ \bar{A}_{P}^{T_P^1}
\end{bmatrix},
\Lambda 
=\begin{bmatrix}
   {\bar{B}} & {{0}} & \cdots  & {{0}}  \\
   \bar{A}\bar{B} & {\bar{B}} & \cdots  & {{0}}  \\
   \vdots & \vdots  & {\ddots} & \vdots  \\
   {{{\bar{A}}}^{T_{P}^{1}-1}}\bar{B} & {{{\bar{A}}}^{T_{P}^{1}-2}}\bar{B} & \cdots  & {\bar{B}} 
\end{bmatrix}.
\label{equ-26} \end{equation}
Substituting \EquRef{equ-25} into \EquRef{equ-18}, we obtain the safety constraint expressed in terms of ${U}^{i}$ as
\begin{equation}\sum_{i\in \mathcal{N}}{A^i U^i - b^i \in \mathcal{K}} , \label{equ-27}\end{equation}
where
\begin{subequations}\begin{align}
{{A}^{i}}&={{G}^{i}}\Lambda , \label{equ-28a}\\ 
{{b}^{i}}&={{H}^{i}}-{{G}^{i}}\Gamma X_{0}^{i} . \label{equ-28b}
\end{align}  \end{subequations}
Thus, the optimization problem in \EquRef{equ-18} can be rearranged as:
\begin{equation}
\mathbf{P}_3^t: \left.
   \begin{aligned}
   \underset{{{U}^{i}},i\in \mathcal{N}}{\mathop{\min }}\,\sum\limits_{i\in \mathcal{N}}{{{\mathcal{F}}^{i}}({{U}^{i}})+{{\mathcal{I}}_{\mathcal{K}}}(\omega )}  \\
   \text{s.t.} \quad \sum\limits_{i\in \mathcal{N}}{({{A}^{i}}{{U}^{i}}-{{b}^{i}})=\omega } ,
   \end{aligned}
\right.
\label{equ-29}\end{equation}
where ${{\mathcal{F}}^{i}}({{U}^{i}})={{F}^{i}}({{U}^{i}},{{X}^{i}})={{F}^{i}}({{U}^{i}},\Gamma X_{0}^{i}+\Lambda {{U}^{i}})$. 
Given that $\Gamma$, $\Lambda$, and $X_{0}^{i}$ are constants, the final ${{\mathcal{F}}^{i}}$ is a function of ${{U}^{i}}$. 
The optimization problem \EquRef{equ-29} is convex and satisfies the Slater condition. 
According to the Lagrange duality theorem, the Lagrangian function of \EquRef{equ-29} is given by:
\begin{equation}
\mathcal{L}=\sum\limits_{i\in \mathcal{N}}{{{\mathcal{F}}^{i}}({{U}^{i}})}+{{\mathcal{I}}_{\mathcal{K}}}(\omega )+\left\langle y,\sum\limits_{i\in \mathcal{N}}{({{A}^{i}}{{U}^{i}}-{{b}^{i}})}-\omega  \right\rangle ,
\label{equ-30}\end{equation}
where $y\in {{R}^{NT_{P}^{1}}}$ represents the dual variable. 
Then, the Lagrangian dual function is given by
\begin{equation}\begin{aligned}
{{d}_{\mathcal{L}}}(y)&=\underset{({{U}^{i}},\omega )}{\mathop{\inf }}\,\mathcal{L}  \\
&=
\underbrace{\underset{{{U}^{i}}}{\mathop{\inf }}\,\left\{ \sum\limits_{i\in \mathcal{N}}{{{\mathcal{F}}^{i}}(}{{U}^{i}})+\left\langle y,{{A}^{i}}{{U}^{i}} \right\rangle  \right\}}_{a} \\
&+
\underbrace{\underset{\omega }{\mathop{\inf }}\,\left\{{{\mathcal{I}}_{\mathcal{K}}}(\omega )-\left\langle y,\omega  \right\rangle \right\}}_{b}
-
\sum\limits_{i\in \mathcal{N}}{\left\langle y,{{b}^{i}} \right\rangle } ,
\end{aligned}\label{equ-31}\end{equation}
In \EquRef{equ-31}, $a$ can be expressed in the form of a convex conjugate \cite{boyd2004convex}. 
For a function $f(x)$, its convex conjugate is defined as
\begin{equation}
   {{f}^{*}}(y):=\underset{x}{\mathop{\sup }}\,\{{{y}^{T}}x-f(x)\}
\label{equ-32}\end{equation}
Thus, we have:
\begin{equation}\begin{aligned}
   a&=-\underset{{{U}^{i}}}{\mathop{\inf }}\,\left\{ -\sum\limits_{i\in \mathcal{N}}{{{\mathcal{F}}^{i}}(}{{U}^{i}})+(-{{y}^{T}}{{A}^{i}}){{U}^{i}} \right\}  \\
   &=-\sum\limits_{i\in \mathcal{N}}{\left\{ \underset{{{U}^{i}}}{\mathop{\inf }}\,{{(-{{A}^{iT}}y)}^{T}}{{U}^{i}}-{{\mathcal{F}}^{i}}({{U}^{i}}) \right\}}  \\
   &=-\sum\limits_{i\in \mathcal{N}}{{{\mathcal{F}}^{i*}}(-{{A}^{iT}}y)} ,
\end{aligned}\label{equ-33}\end{equation}
As for $b$ in \EquRef{equ-31}, considering the presence of the indicator function ${\mathcal{I}}_{\mathcal{K}}$, if $\omega$ is not in $\mathcal{K}$, this term results in infinity. 
Therefore, it must hold that $\omega \in \mathcal{K}$. 
Hence,
\begin{equation}\begin{aligned}
   b&=\underset{\omega }{\mathop{\inf }}\,\{{{\mathcal{I}}_{\mathcal{K}}}(\omega )-\left\langle y,\omega  \right\rangle \}  \\
   &=\underset{\omega \in \mathcal{K}}{\mathop{\inf }}\,\{\underbrace{{{\mathcal{I}}_{\mathcal{K}}}(\omega )}_{0}-\left\langle y,\omega  \right\rangle \}  \\
   &=-{{\mathcal{I}}_{{{\mathcal{K}}^{{}^\circ }}}}(y) . \\
\end{aligned}\label{equ-34}\end{equation}
The last equality comes from \cite{banjac2019decentralized}, where ${{\mathcal{K}}^{{}^\circ }}$ denotes the polar cone of $\mathcal{K}$. 
Consequently, the Lagrangian dual problem of \EquRef{equ-29} can be formulated as
\begin{equation}
\!\!\! \mathbf{P}_4^t : \left.
\begin{aligned}
&\underset{y}{\mathop{\max }}\,{{d}_{\mathcal{L}}}(y) \\
=&\underset{y}{\mathop{\max }}\!\left\{ a+b-\sum\limits_{i\in \mathcal{N}}{\left\langle y,{{b}^{i}} \right\rangle } \right\}  \\
=&\underset{y}{\mathop{\max }}\!\left\{ \!\sum\limits_{i\in \mathcal{N}}{\left[ -\left\langle y,{{b}^{i}} \right\rangle \!-\!{{\mathcal{F}}^{i*}}(-{{A}^{iT}}y) \right]}\!-\!{{\mathcal{I}}_{{{\mathcal{K}}^{{}^\circ }}}}(y) \! \right\} \\ 
=&\underset{y}{\mathop{\max }}\!\left\{ \!\sum\limits_{i\in \mathcal{N}}{\left[ -\left\langle y,{{b}^{i}} \right\rangle \!-\!{{\mathcal{F}}^{i*}}(-{{A}^{iT}}y)\!-\!{{\mathcal{I}}_{{{\mathcal{K}}^{{}^\circ }}}}(y) \right]} \! \right\} . \\ 
\end{aligned}
\right.
\label{equ-35}\end{equation}

The equality in \EquRef{equ-35} holds due to ${{\mathcal{I}}_{{{\mathcal{K}}^{{}^\circ }}}}(y)=C\times {{\mathcal{I}}_{{{\mathcal{K}}^{{}^\circ }}}}(y)$, where $C$ is an arbitrary constant.

\subsection{Decomposition and solve the dual problem}\label{sec-3-2}

The optimization problem of \EquRef{equ-35} is a global variable consensus problem, 
characterized by the fact that each vehicle (node) has a different objective function, 
but they collectively optimize variable $y$ to maxmize the extreme value sum of all objective functions, 
and the collectively optimized variable will eventually become ``consensus''.
The basic method for solving this problem using ADMM was proposed by Byod in \cite{boyd2011distributed}. 
In \cite{mateos2010distributed}, the authors proposed two distributed algorithms based on ADMM to solve the global variable consensus problem, 
and we adopts the second one. 
Firstly, each vehicle's objective function is decomposed into two parts, i.e., $d_{\mathcal{L}}^{i,1}$ and $d_{\mathcal{L}}^{i,2}$,
\begin{equation} 
{{d}_{\mathcal{L}}}(y)=
\sum\limits_{i \in \mathcal{N}}{ 
   \underbrace{-[\left\langle y,{{b}^{i}} \right\rangle + {{\mathcal{F}}^{i*}}(-{{A}^{iT}}y)]}_{d_{\mathcal{L}}^{i,1}}
   +
   \underbrace{[-{{\mathcal{I}}_{{{\mathcal{K}}^{{}^\circ }}}}(y)]}_{d_{\mathcal{L}}^{i,2}} 
}.
\label{equ-36}\end{equation}

Then, we adopt \AlgRef{algo-1} to solve the problem is used for solving. 
The algorithm requires two inputs, i.e., $\sigma$ and $\rho$.
They are two hyperparameters, similar to step size in optimization algorithms.
In line 1, variable $k$ is used to mark the number of iterations and each vehicle initializes local variables, i.e. $y^{i,k}$, $p^{i,k}$, and $s^{i,k}$. 
They are typically initialized as zero vectors.
$y^{i,k}$ is the local optimization variables of vehicle $i$ at time step $k$.
As the iterations progress, $y^{i,k}$ of all users will gradually converge to the same vector, forming a consensus. 
At the beginning of each iteration, vehicles need to exchange local optimization variables $y^{i,k}$ with each other (line 3). 
Then, vehicles need to calculate $p^{i,k+1}$ and $s^{i,k+1}$ in line 4 and 5, respectively.
After that, vehicles need to solve two optimization problem in line 6 and 7.
Since there are time constraints for solving the trajectory planning problem, the termination condition in line 8 will be set as a fixed number of iterations in the experimental section. 
The goal of the algorithm is to obtain a consensus solution for \EquRef{equ-35}, so the convergence of the algorithm can be judged by checking whether the variance of the local variables ${{y}^{i,k}} $ of all vehicles converges \cite{huang2023decentralized}.

\input{misc/algo01.tex}

When using \AlgRef{algo-1}, each vehicle can only obtain the local dual variable $y^{i,k}$ and cannot directly obtain the original problem's optimization variable ${{U}^{i}}$. 
In addition, the optimization problems in line 6 and 7 are complicated and difficult to solve.
To solve quickly and obtain the solution to the original problem, \textbf{Proposition 2} from \cite{banjac2019decentralized} is used. 
This proposition states that when solving the optimization problem in line 6, the original problem's optimization variable $U^{i}$ for this round of iteration can be simultaneously obtained. 
Furthermore, as the iterations progress, $U^{i}$ will gradually converge to the optimal solution of the original problem. 
To use the conclusion, we consolidate the first three terms of the optimization objective in line 6 of \AlgRef{algo-1} and eliminate the irrelevant constant terms and we get
\begin{subequations}\begin{align}
&\underset{y}{\arg\min}
\left\{ 
   d_{\mathcal{L}}^{i,1}(y) + \frac{\sigma + 2 \rho d_{i}}{2} \left\| y - \frac{r^{i,k+1}}{\sigma + 2 \rho d_i} \right\|^2 
\right\} , \label{equ-37a} \\
&
\begin{aligned}
r^{i,k+1} &= \sigma z^{i,k} + \rho \sum_{j \in \mathcal{N}_i} (y^{i,k}+y^{j,k}) \\ &-(b^{i} + p^{i,k+1} + s^{i,k+1}),
\end{aligned} \label{equ-37b}
\end{align}\end{subequations}
where ${{d}_{i}}$ represents the degree of vehicle $i$ in the fully connected graph formed by the vehicular network.
Therefore, for $\forall i\in \mathcal{N}$, ${{d}_{i}}=(N-1)$. 
The term $b_i$ is derived from \EquRef{equ-28b}. 
According to \cite{banjac2019decentralized}, the solution to problem \EquRef{equ-37a} is given by
\begin{subequations}\begin{align}
  y&=\frac{{{A}^{i}}{{U}^{i,k+1}}+{{r}^{i,k+1}}}{\sigma +2\rho {{d}_{i}}} , \label{equ-38a} \\ 
  U^{i,k+1} \! &= \underset{U}{\arg\min}
 \left\{ 
   \mathcal{F}^i(U)
   \!+\!
   \frac{\left\| A^i U \!+\! r^{i,k+1} \right\|^2}{2(\sigma + 2 \rho d_i)} 
 \right\} . \label{equ-38b}
\end{align}\end{subequations}
where $A^i$ and $\mathcal{F}^i$ are from \EquRef{equ-28a} and \EquRef{equ-29}, respectively. 
Next, expanding the optimization problem in \EquRef{equ-38b}, we get
\begin{equation}
\mathbf{P}_5^t: \left.
\begin{aligned}
   \underset{U^i}{\min} \quad & U^{iT} (Q \otimes E^{T_{P}^{1}} + \frac{1}{2(\sigma + 2 \rho d_i)} A^{iT} A^i) U^i \\
   & + 2 \frac{1}{2(\sigma +2 \rho d_i)} (A^{iT} r)^{T} U^i \\
   \text{s.t.} \quad & U_l^i \preceq {\begin{bmatrix} E^{2T_P^1}  \\ M_X^i \Lambda \end{bmatrix}} U^i \preceq U_u^i .
\end{aligned} 
\right.
\label{equ-39}
\end{equation}
Since the quadratic term in the objective function is positive definite, this problem is a standard convex quadratic programming problem that can be quickly solved. 
Here, $M_X^i$, $U_u^i$, and $U_l^i$ represent constraints on $U^i$ derived from the non-coupling constraints of the dynamic constraint \EquRef{equ-25} applied in \EquRef{equ-8}, with their values given by
\begin{subequations}\begin{align}
& U_{l}^{i}= 
\begin{bmatrix} 
   \underline{U} \otimes 1^{T_P^1} \\
   \begin{bmatrix}
   (L_1 + L_2)+(\mathcal{S}_i - 1) L_f \\ \left(L_1 - s_0^i\right) / T_s
   \end{bmatrix} -M_{X}^{i} \Gamma X_{0}^{i}  \\
\end{bmatrix} , \\ %\in R^{3 T_P^1 + 2} \\ 
& U_u^i = 
\begin{bmatrix}
   \overline{U} \otimes 1^{T_P^1}  \\
   \begin{bmatrix}
   (L_1 + L_2) + \mathcal{S}_i L_f  \\  \left(L_1 + L_2 - s_0^i\right) / T_s
   \end{bmatrix} - M_X^i \Gamma X_0^i  \\
\end{bmatrix} , \\ %\in R^{3T_P^1 + 2} \\ 
&M_{X}^{i}=\left[
\setlength{\arraycolsep}{3pt}
\begin{array}{lr}
      1^{1 \times (T_P^1-1)} \otimes 0^{1 \times 3} 
      &
      \begin{bmatrix} 1 \, 0 \, 0 \end{bmatrix} \\
      % \hdashline
      1^{1 \times t_i^M} \otimes \begin{bmatrix} 0 \, 1 \, 0 \end{bmatrix} 
      & 
      1^{1 \times (T_P^1 - t_i^M)} \otimes 0^{1 \times 3}
\end{array}
\right] .
\end{align}\end{subequations}

The optimization problem in line 7 of \AlgRef{algo-1} is relatively simple. 
Considering that $d_{\mathcal{L}}^{i,2}$ is the characteristic function on the set $\mathcal{K}^{\circ}$, the solution to this optimization problem must lie within the set $\mathcal{K}^{\circ}$. 
Since the optimization variable is $z$, by completing the square of the objective function with respect to $z$ and removing irrelevant constant terms, we can obtain
\begin{equation}
\begin{aligned}
z^{i,k+1} 
&= \underset{z \in \mathcal{K}^\circ}{\arg\min} \left\{ 
   \frac{\sigma}{2}{\left\| z - y^{i,k+1} \right\|^2} - \left\langle z, s^{i,k+1} \right\rangle  
\right\} \\
&= \underset{z \in \mathcal{K}^\circ}{\arg\min} \left\{ 
   \left\| z - \left( y^{i,k+1} + \frac{s^{i,k+1}}{\sigma} \right) \right\|^2 
\right\}.
\end{aligned}
\end{equation}

Since $\mathcal{K}^\circ$ is a convex cone formed by negative limits, truncating each element of the vector $y^{i,k+1} + \frac{s^{i,k+1}}{\sigma}$ on the negative real axis yields the solution
\begin{equation}
z^{i,k+1}= \min \left\{ y^{i,k+1} + \frac{s^{i,k+1}}{\sigma}, 0 \right\}.
\label{equ-42}\end{equation}

In summary, the entire trajectory planning algorithm is summarized in \AlgRef{algo-2}. The vehicle-to-vehicle communication in lines 3 and 6 can be implemented using V2X technology to achieve multi-vehicle collaboration.

\input{misc/algo02.tex}

%%%%%%%%%%%%%%%%%%%%%%%%%%%%%%%%%%%%%%%%%%%%%%%%%%%%%%%%%%%%%%%%%%%%%%%%%%%%%%%%%%%%%%%%%%%%%%%%%%%%%%%%%%%%%%%%%%%%%%
\section{Distributed Cooperative Iterative MPC}\label{sec-4}

The optimization problem of \EquRef{equ-14} is different from the trajectory planning problem in \SecRef{section-2-1}. 
During the entire ramp merging process, the latter only needs to be solved once at the beginning of the merging process, with lower requirements for solving speed. 
However, MPC requires solving the problem at each time step, so how to quickly solve this problem is crucial. 
As described in \SecRef{sec-2-2}, due to the presence of safety constraints in problem \EquRef{equ-14c}, 
the cooperative control is introduced but the optimization problems of each vehicle are coupled and non-convex, 
which generates a computationally intensive task and makes it difficult to solve.
Therefore, a method for rapidly and parallelly solving this problem is proposed in this section, i.e., distributed cooperative iterative MPC.
\SecRef{sec-4-1} introduces how to decouple the problem and distribute each non-convex subproblem to local vehicles for parallel computation. 
And \SecRef{sec-4-2} explains how to convexify the non-convex subproblems and solve them quickly.

\subsection{Decomposition of the Coupled Control Problem}\label{sec-4-1}

Next, for clarity of exposition, the subsequent $T_{P}^{2}$ state variables obtained by solving the optimization problem for vehicle $i$ at time step $k$ are denoted as $\bar{X}_{(k+l,k)}^{i}$, where $l\in [1,T_{P}^{2}]$. 
These variables concatenated into a single column vector to form a trajectory are commonly referred to as a nominal trajectory, denoted as $\bar{X}_k^i=[\bar{X}_{(k+1,k)}^{iT}, \bar{X}_{(k+2,k)}^{iT}, \cdots ,\bar{X}_{(k+T_{P}^{2},k)}^{iT}]^T$. 
Since the nominal trajectory contains information about the future driving trajectory, vehicles can first transmit nominal trajectories to each other through V2X and then establish local safety constraints and MPC problems based on nominal trajectories, finally solving the problem and performing collision avoidance in advance. 
In this way, the centralized problem \EquRef{equ-14} can be solved in a distributed manner by solving the following optimization problem $\mathbf{P}_2^c$ on each vehicle to obtain a suboptimal solution to the original problem.
\begin{subequations}\begin{align}
 \underset{\substack{U_{\left( k+l,k \right)}^{i}, \\ l \in \left[ 0, T_P^2-1 \right]}}{\min} \; & \sum_{l=1}^{T_P^2 - 1} \left\| M_f \left( X_{(k+l+1,k)^i} - X_{(k+l,k)^i} \right) \right\|^2 \nonumber \\
& \!\!\!\!\!\! + \sum_{l=1}^{T_P^2} \left\| X^i_{(k+l,k)} - \bar{X}^i_{k+l,k} \right\|^2_{Q_X} 
\!\!\!
+ \left\| U^i_{(k+l-1,k)} \right\|^2_{Q_U} \nonumber \\
\mathbf{P}_2^c: \;\;\; \text{s.t.} \;
& \underline{U} \le U_{\left( k+l,k \right)}^{i}\le \overline{U} , \\
& X_{(k+l+1,k)}^{i} = f_d \left( X_{(k+l,k)}^{i},U_{(k+l,k)}^{i} \right) , \label{equ-43b}\\
& D_s \le D \left( X_{(k+l,k)}^{i},X_{(k+l,k)}^{j} \right) , \label{equ-43c} \\
& \forall l \in \left[ l, T_P^2 \right], j \in \mathcal{N}_i . \nonumber
\end{align}\label{equ-43}\end{subequations}

For problem \EquRef{equ-43}, its objective function is a positive definite quadratic form.
However, \EquRef{equ-43b} and \EquRef{equ-43c} are both nonlinear and non-convex constraint conditions, which will lead to difficulties in solving.
For the nonlinear time-varying vehicle kinematic model constraint in \EquRef{equ-43b}, many works use Taylor series expansions for first-order approximations near the reference trajectory. 
However, due to the presence of safety constraints in \EquRef{equ-43c}, the actual driving trajectory that vehicles can avoid collisions will differ significantly from the reference trajectory. 
In this case, if a first-order Taylor approximation is still applied to the $f_d$ near the reference trajectory, the error will be significant.
Therefore, at time $k$, we use the nominal trajectory obtained at time $k-1$ to perform a first-approximation of the constraint in \EquRef{equ-43b}.
After solving, $\bar{X}_{k}^{i}$ is obtained, and this trajectory is then extended by a sampling time and used as the nominal trajectory at time $k+1$ for the next iteration.

Then, for clarity of presentation, the following definitions are made:
\begin{equation}
\begin{aligned}
X_k^i = \begin{bmatrix}
X_{(k+1,k)}^{iT} & X_{(k+2,k)}^{iT} & \ldots & X_{(k+T_P^2,k)}^{iT}
\end{bmatrix}^T \in R^{4T_P^2} , \\
U_k^i = \begin{bmatrix}
U_{(k,k)}^{iT} & U_{(k+1,k)}^{iT} & \ldots & U_{(k+T_P^2,k)}^{iT}
\end{bmatrix}^T \in R^{2T_P^2} .
\end{aligned}
\label{equ-44}\end{equation}
For the discretized nonlinear system function $f_d$ in \EquRef{equ-13}, first-order approximation is taken at the nominal trajectory $\bar{X}_{(k+l,k)}^i$, $\bar{U}_{(k+l,k)}^i$ as
\begin{equation}
\begin{aligned}
  & f_{d}(X_{(k+l,k)}^{i},U_{(k+l,k)}^{i}) \\
% = & T_{S}f(X_{(k+l,k)}^{i},U_{(k+l,k)}^{i}) + X_{(k+l,k)}^{i} \\
% \approx & T_{S}f(\bar{X}_{(k+l,k)}^{i},\bar{U}_{(k+l,k)}^{i}) + \bar{X}_{(k+l,k)}^{i} \\
% + &
% \left[ T_{S} \mathcal{J}^f_X \left( \bar{X}^i_{(k+l,k)}, \bar{U}^i_{(k+l,k)} \right)
% \!+\! E \right] \left( X_{(k+l,k)}^i - \bar{X}_{(k+l,k)}^i \right) \\
% + & \left[ T_{S} \mathcal{J}^f_U \left( \bar{X}^i_{(k+l,k)}, \bar{U}^i_{(k+l,k)} \right)
% \right] \left( \bar{U}_{(k+l,k)}^i - \bar{U}_{(k+l,k)}^i \right) \\
\approx & A_{(k+l,k)}^{i}X_{(k+l,k)}^{i} + B_{(k+l,k)}^{i}U_{(k+l,k)}^{i} + G_{(k+l,k)}^{i},
\end{aligned}
\label{equ-45}\end{equation}
where
\begin{equation}
\begin{aligned}
A_{(k+l,k)}^i &=  T_{S} \mathcal{J}^f_X \left( \bar{X}^i_{(k+l,k)}, \bar{U}^i_{(k+l,k)} \right)
+ E , \\
B_{(k+l,k)}^{i} &= T_{S} \mathcal{J}^f_U , \\
G_{(k+l,k)}^{i}
& = T_S [ 
   f(\bar{X}_{(k+l,k)}^{i},\bar{U}_{(k+l,k)}^{i}) \\
   &- \mathcal{J}^f_X \left( \bar{X}^i_{(k+l,k)}, \bar{U}^i_{(k+l,k)} \right) \bar{X}^i_{(k+l,k)}\\
   &- \mathcal{J}^f_U \left( \bar{X}^i_{(k+l,k)}, \bar{U}^i_{(k+l,k)} \right) \bar{U}^i_{(k+l,k)}
].
\end{aligned}\label{equ-46}
\end{equation}
Similar to \EquRef{equ-24}, the constraints in \EquRef{equ-43b} can be organized as linear constraints on the initial state $X_{(k,k)}^{i}$ and $X^{i}$ with $U^{i}$ as
\begin{equation}
X_k^i = A_{k}^{i}X_{(k,k)}^{i} + B_{k}^{i}U_{k}^{i} + G_{k}^{i},
\label{equ-47}\end{equation}
and the definitions of them are shown in \EquRef{equ-48}.
\begin{figure*}[hbtp]
\begin{equation}\begin{aligned}
&A_k^i = \begin{bmatrix}
   A_{(k,k)}^i \\
   A_{(k+1,k)}^i A_{(k,k)}^i \\
   \vdots \\
   \prod_{l=0}^{T_P^2-1} A_{(k+l,k)}^i
\end{bmatrix},
G_k^i = \begin{bmatrix}
   G_{(k,k)}^i \\
   A_{(k+1,k)}^i G_{(k,k)}^i + G_{(k+1,k)}^i \\
   \vdots \\
   \sum_{\epsilon=0}^{T_P^2-2}{\left( \prod_{l=\epsilon +1}^{T_P^2-1}{A_{(k+l,k)}^i} \right) G_{(k+\epsilon ,k)}^i + G_{(k + T_P^2 - 1,k)}^{i}} \\
\end{bmatrix},\\
&B_{k}^{i} = 
\begin{bmatrix}
   B_{(k,k)}^i & \cdots & \cdots & 0 \\
   A_{(k+1,k)}^{i}B_{(k,k)}^{i} & B_{(k+1,k)}^{i} & \cdots & 0 \\
   \vdots & \vdots & \ddots & \vdots \\
   \prod_{l=1}^{T_P^2-1} A_{(k+l,k)}^i B_{(k,k)}^i
   & \prod_{l=2}^{T_P^2-1} A_{(k+l,k)}^i B_{(k+1,k)}^i
   & \cdots 
   & B_{(k + T_P^2 - 1,k)}^i \\
\end{bmatrix}
\end{aligned}\label{equ-48}\end{equation}
\end{figure*}
% where $A^i \in R^{4T_{P}^2 \times 4}$, $B^i \in R^{4T_P^2 \times 2T_P^2}$, and $G^i \in R^{4T_P^2}$. 
However, it should be noted that, due to the time-varying nature of the kinematic, the matrices $A_{(k+l,k)}^i$, $B_{(k+l,k)}^i$, and $G_{(k+l,k)}^i$ used to compute the matrices in \EquRef{equ-47} will vary with different $l$.
Finally, we can organize \EquRef{equ-43} into a form of linear kinematic constraints as
\begin{subequations}
\begin{align}
\underset{X_k^i , U_k^i}{\min}\quad &
\left\| X_k^i - \tilde{X}_k^i \right\|_{\bar{Q}_X}^2 
+ \left\| U_k^i \right\|_{\bar{Q}_U}^2 + 
\left\| \bar{M}_f X_k^i \right\|^2 \nonumber \\
\mathbf{P}_3^c: \;\;\;\;\; \text{s.t.} \quad &  E^{T_{p}^{2}} \otimes \underline{U} \leq U_k^i \leq E^{T_p^2} \bar{U} \\
& X_k^i = A_k^i X_{(k,k)}^i + B_k^i U_k^i + G_k^i \\
& E^{T_p^2} \otimes D_s \leq \mathcal{D}(X_k^i , \bar{X}_k^j), j \in \mathcal{N}_i \label{equ-49c}
\end{align}
\label{equ-49}\end{subequations}
where the function 
$\mathcal{D}(X^i_k, \bar{X}^j_k): R^{4T_P^2 \times 4T_P^2} \to R^{4T_P^2}$
is a function formed by concatenating the function $D$ column-wise as
\begin{equation}
\mathcal{D}(X^i_k, \bar{X}^j_k) = 
\begin{bmatrix}
D(X_{(k+1,k)}^i, X_{(k+1,k)}^j) \\
D(X_{(k+2,k)}^i, X_{(k+2,k)}^j) \\
\ldots \\
D(X_{(k+T_P^2,k)}^i, X_{(k+T_P^2,k)}^j)
\end{bmatrix}^T .
\label{equ-50}\end{equation}
The definitions of $\bar{Q}_X$, $\bar{Q}_U$, and $\bar{M}_f$ are given by
\begin{equation}
\begin{aligned}
{\bar{Q}_{X}} &= \left[ \begin{array}{c:c}
E^{T_P^2-1} \otimes Q_X & 0  \\
\hdashline
0 & k_X Q_X
\end{array} \right] , \\
{\bar{Q}_U} &= \left[ \begin{array}{c:c}
E^{T_{P}^{2}-1} \otimes Q_{U} & 0  \\
\hdashline
0 & k_U Q_U
\end{array} \right] , \\
\bar{M}_f &= \begin{bmatrix}
-M_f & M_f & \space & \space & \space & \space \\
\space & -M_f & M_f & \space & \space & \space \\
\space & \space & \space & \ddots & \space & \space \\
\space & \space & \space & \space & -M_f & M_f
\end{bmatrix} ,
\end{aligned}
\label{equ-51}\end{equation}
where $k_{X}$ and $k_{U}$ are used to constrain the last point of the trajectory in model predictive control, and the elements not labeled in $\bar{M}_{f}$ are zeros.

\subsection{Convex Reformulation of Subproblem}\label{sec-4-2}

\begin{figure*}[b]
\centering 
\includegraphics[scale=0.8]{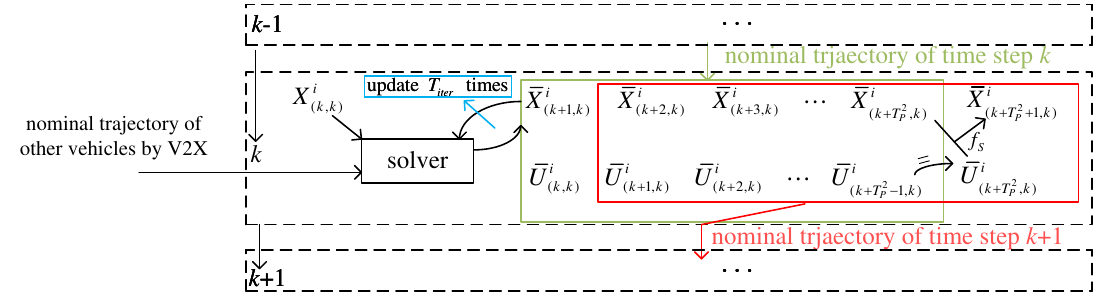}
\centering
\caption{The variation process of nominal trajectories in DCIMPC}\label{DCIMPC}
\end{figure*}

In this subsection, we will explains how to reformulate problem \EquRef{equ-49} into a convex QP problem. 
For simplify, we define 
\begin{equation}\begin{aligned}
& \tilde{D}(X_{(k+l,k)}^{i},X_{(k+l,k)}^{j},p,q) \\ 
= & \min \left\{ \dot{D}(X_{(k+l,k)}^{i},X_{(k+l,k)}^{j},p,q)-{{D}_{S}},0 \right\}.
\end{aligned}\label{equ-52}\end{equation}
Satisfying the safety constraint \EquRef{equ-49c} is equivalent to $\forall i \in \mathcal{N}$, $j \in \mathcal{N}_i$, $ l \in \left\| 1, T_P^2 \right\|$ and $p,q \in \{-1, 1\}$,
\begin{equation}
\tilde{D}(X_{(k+l,k)}^{i},X_{(k+l,k)}^{j},p,q)^2 \le 0.
\label{equ-53}\end{equation}
Then, we define
\begin{equation}
\mathcal{T}_{k,l}^{i,j}:= \begin{bmatrix}
\tilde{D}(X_{(k+l,k)}^i , X_{(k+l,k)}^j , 1 , 1)  \\
\tilde{D}(X_{(k+l,k)}^i , X_{(k+l,k)}^j , 1 , -1)  \\
\tilde{D}(X_{(k+l,k)}^i , X_{(k+l,k)}^j , -1 , 1)  \\
\tilde{D}(X_{(k+l,k)}^i , X_{(k+l,k)}^j , -1, -1)
\end{bmatrix}.
\label{equ-54}\end{equation}
Therefore, the safety constraint for vehicle $i$ with respect to neighboring vehicle $j$ at time $k$ is equivalent to
\begin{equation}
\begin{aligned}
{{\left\| \mathcal{T}_{k}^{i,j} \right\|}^{2}}
=&
\left\| \begin{bmatrix}
   \mathcal{T}_{k,1}^{i,j}  \\
   \mathcal{T}_{k,2}^{i,j}  \\
   \vdots   \\
   \mathcal{T}_{k,T_{P}^{2}}^{i,j}
\end{bmatrix} \right\|^2  \\
=&
\sum_{l=1}^{T_P^2}
\sum_{p,q \in \{1,-1\}} \!\!\!\!\!\!\!
\tilde{D}(X_{(k+l,k)}^i , X_{(k+l,k)}^j , p , q)^2
\le 0  . \\
\end{aligned}
\label{equ-55}\end{equation}
Next, this term is used as a soft constraint and included in the objective function. Thus, the objective function becomes
\begin{equation}\begin{aligned}
\bar{J}^i
&=
\left\| X_k^i - \tilde{X}_k^i \right\|_{\bar{Q}_X}^2
+
\left\| U_k^i \right\|_{\bar{Q}_U}^2\\
&+
\left\| \bar{M}_f X_k^i \right\|^2
+
\alpha \sum_{j \in \mathcal{N}_i} \left\| \mathcal{T}_k^{i,j} \right\|^2 ,
\end{aligned}\label{equ-56}\end{equation}
where $\alpha$ is used to control the weight of safety constraint.
However, this term in \EquRef{equ-56} is still a non-convex function of $X_{k}^{i}$, and it needs to be handled next. 
First, for the function inside the norm operation, i.e., $\mathcal{T}_{k}^{i,j}$, a first-order Taylor approximation is taken at the nominal trajectory and the constant term is discarded, resulting in
\begin{equation}
\begin{aligned}
\left\| \mathcal{T}_k^{i,j} \right\|^2
&\approx 
\left\| 
   \mathcal{T}_{k}^{i,j}\left( \bar{X}_{k}^{i},\bar{X}_{k}^{j} \right)
   +
   \left. \frac{\partial \mathcal{T}_k^{i,j}}{\partial X_k^i} \right|_{\bar{X}_k^i,\bar{X}_k^j}
   \left( X_k^i - \bar{X}_k^i \right) 
\right\|^2  \\
&=
\Bigg\| 
   \underbrace{
      {\left. \frac{\partial \mathcal{T}_k^{i,j}}{\partial X_k^i} \right|}
      _
      {\bar{X}_k^i,\bar{X}_k^j}
   }_{k^i} X_k^i
   \\&+
   \underbrace{
      \left[ 
         \mathcal{T}_k^{i,j}\left( \bar{X}_k^i,\bar{X}_k^j \right)
         -
         {\left. \frac{\partial \mathcal{T}_k^{i,j}}{\partial X_k^i} \right|}
         _
         {\bar{X}_k^i,\bar{X}_k^j}
         \bar{X}_k^i 
      \right]
   }_{b^i} 
\Bigg\| ^2  \\
&=
\left\| k^i X_k^i + b_i \right\|^2  \\
&=
X_k^{iT} (k^{iT} k^i) X_k^i + 2(k^{iT} b^i)^T X_k^i  
\end{aligned}
\label{equ-57}\end{equation}
Since \EquRef{equ-57} is a quadratic form in terms of $X_{k}^{i}$, and ${{k}^{iT}}{{k}^{i}}$ is a positive semi-definite matrix, the objective function \EquRef{equ-56} becomes a quadratic convex function at this point, transforming the optimization problem into a convex quadratic programming problem. 
Next, by incorporating the system constraints into the entire optimization problem and eliminating $X_{k}^{i}$ while removing irrelevant constant terms, the simplification of the first term of \EquRef{equ-56} results in
\begin{equation}\begin{aligned}
& \underset{X_k^i,U_k^i} \min \left\| X_k^i - \tilde{X}_k^i \right\|_{\bar{Q}_X}^2 \\
=& \underset{X_k^i,U_k^i} \min \left[ {\left( X_{k}^{i}-\tilde{X}_{k}^{i} \right)}^T \bar{Q}_X \left( X_k^i - \tilde{X}_k^i \right) \right]\\
=& \underset{X_k^i,U_k^i} \min \left[ X_k^{iT} \bar{Q}_X X_k^i - 2\tilde{X}_k^i \bar{Q}_X X_k^i \right]\\
=& \underset{U_k^i} \min \; \bigg \{ U_k^{iT} \left( B_k^{iT} \bar{Q}_X B_k^i \right) U_k^i \\
+&
2\left[ B_k^{iT} \bar{Q}_X \left( A_k^i X_{(k,k)}^i + G_k^i - \tilde{X}_k^i \right) \right]^T
U_k^i \bigg\} .
\end{aligned}\label{equ-58}\end{equation}
Then, the simplification of the second and third terms of \EquRef{equ-56} results in
\begin{equation}\begin{aligned}
& \underset{X_k^i,U_k^i} \min \left\| U_k^i \right\|_{\bar{Q}_U}^2 + 
\left\| \bar{M}_f X_k^i \right\|^2 \\
=& \underset{X_k^i,U_k^i} \min \left\| U_k^i \right\|_{\bar{Q}_U}^2 + \left\| \bar{M}_f X_k^i \right\|^2 \\
=& \underset{U_k^i} \min \; 
U_k^{iT} \left( B_k^{iT} \bar{M}_f^T \bar{M}_f B_k^i + \bar{Q}_U \right) U_k^i 
\\
+& 2 \left[ B_k^{iT} \bar{M}_f^T \bar{M}_f \left( A_k^i X_k^i + G_k^i \right) \right]^T U_k^i .
\end{aligned}\label{equ-59}\end{equation}
The simplification of the fourth term leads to
\begin{equation}\begin{aligned}
\underset{X_k^i} \min \; & \alpha \sum_{j\in {\mathcal{N}_i}} 
X_k^{iT} (k^{iT} k^i) X_k^i + 2 (k^{iT} b^i)^T X_k^i \\
= \underset{U_k^i} \min \; & \alpha \Bigg[ 
   U_k^{iT} B_k^{iT}
   \sum_{j \in \mathcal{N}_i} \left( k^{iT} k^i \right)
   B_k^i U_k^i 
   \\ &+
   2 \left( 
      B_k^{iT} \sum_{j \in \mathcal{N}_i} k^{iT} b^i 
   \right)^T
   U_k^i 
\Bigg] .
\end{aligned}\label{equ-60}\end{equation}
Finally, we obtain the following standard quadratic programming problem $\mathbf{P}_4^c$, 
\begin{equation}\label{equ-61}
\mathbf{P}_4^c: \left.
\begin{aligned}
\underset{U_k^i} \min \;\; & J^i = \underset{U_k^i} \min \;\; U_k^{iT} \mathcal{P} U_k^i + 2 \mathcal{Q}^T U_k^i \\
\text{s.t.} \;\; & E^{T_p^2} \otimes \underline{U} \le U_k^i \le E^{T_p^2} \otimes \overline{U} ,
\end{aligned}
\right.
\end{equation}
where 
\begin{equation}\label{equ-62}\begin{aligned}
& \mathcal{P} = 
B_k^{iT} \left( \bar{Q}_X + \alpha \sum_{j \in \mathcal{N}_i} k^{iT} k^{i} + \bar{M}_f^T \bar{M}_f \right) B_k^i + \bar{Q}_U 
, \\
& \mathcal{Q} = B_{k}^{iT} 
\Bigg[ 
\alpha \sum_{j \in \mathcal{N}_i} k^{iT} b^i - \bar{Q}_X \tilde{X}_{(k,k)}^i\\
&+
\bigg( 
   \bar{Q}_X + \alpha \sum_{j \in \mathcal{N}_i }{k^{iT} k^i} + \bar{M}_f^T \bar{M}_f
\bigg)
\left( 
   A_k^i X_{(k,k)}^i + G_k^i 
\right)
\Bigg] .
\end{aligned}\end{equation}

In summary, the solution steps for the problem in \EquRef{equ-61} are summarized as \AlgRef{algo-3}. 
It should be note that, a hyperparameter $T_{iter}$ is introduced, which represents the number of iterations for MPC at each time step. 
As mentioned \SecRef{sec-4-1}, the precision of \EquRef{equ-61} relys on nominal trajectories. 
To further improve accuracy, an iterative model predictive control approach is introduced. 
This algorithm is more suitable for time-varying kinematic models and safety constraints. 
At each time step $k$, vehicle $i$ assigns the number of iterations to a variable $C$ (Line 3). 
Then, the iterative model predictive control is initiated. 
At the beginning of the iteration, vehicles exchange their nominal trajectories using V2X communication (Lines 5 and 6). 
The system constraints are linearized based on the vehicle's nominal trajectory by \EquRef{equ-47}, and then the Jacobian matrix $k^j$ and vector $b^j$ are computed based on other vehicles' nominal trajectories by \EquRef{equ-57}. 
These values are then used to construct the optimization problem of \EquRef{equ-62} (Line 7), and the OSQP solver is used to solve it, with the nominal trajectory as the starting point for it (Line 8) and solved (Line 9) \cite{stellato2020osqp}. 
Next, based on the obtained $U_{(k+l,k)}^{i*}, l\in [0,T_{P}^{2}-1]$, a new nominal trajectory is generated (Line 10), and the constant $C$ is decremented by 1 (Line 11). 
When $C$ reaches 0, the current iteration ends, the first control variable for the final result is executed (Line 13), and the nominal trajectory for the next time step is generated (Line 14).

The generation and utilization of nominal trajectories during the solving process are illustrated in \FigRef{DCIMPC}. 
At time $k$, vehicle $i$ uses the nominal trajectories left from time $k-1$ as the initial nominal trajectory (marked by the green rectangle in the \FigRef{DCIMPC}). 
These trajectories are updated at each iteration and extended by a sampling time to generate the nominal trajectory for time $k+1$ for the next use. 
The extension method is shown in the red box in \FigRef{DCIMPC}, where the control variable at $k+T_{P}^{2}$ is set to the control variable at $k+T_{P}^{2}-1$, then $U_{(k+T_{P}^{2},k)}^{i}$ and $X_{(k+T_{P}^{2},k)}^{i}$ are input into the system function \EquRef{equ-10} to compute the state at $k+T_{P}^{2}+1$, which is combined with previous results to generate the initial nominal trajectory for the next iteration , as shown in Line 15 of \AlgRef{algo-3}. 
Finally, $k$ is incremented by 1, and the process returns to Line 4 for the next control iteration.

\input{misc/algo03.tex}

%%%%%%%%%%%%%%%%%%%%%%%%%%%%%%%%%%%%%%%%%%%%%%%%%%%%%%%%%%%%%%%%%%%%%%%%%%%%%%%%%%%%%%%%%%%%%%%%%%%%%%%%%%%%%%%%%%%%%%
\section{Simulation Results}\label{sec-5}

In the first subsection of this part, we first generate a two-dimensional reference trajectory using the trajectory planning algorithm \AlgRef{algo-2} proposed in \SecRef{sec-3}, and then conduct overall ramp merging simulations based on this trajectory using the DCIMPC algorithm \AlgRef{algo-3} introduced in \SecRef{sec-4}. 
In the second subsection, further simulation experiments are conducted on the DCIMPC.

The simulations in this work are based on the Python and the open-source symbolic computation library CasADi \cite{andersson2019casadi}. 
CasADi library is capable of computing algebraic differentials and integrates various optimization solvers such as OSQP, IPOPT, and others \cite{biegler2009large, stellato2020osqp}. 
The hardware platform used is the AMD Ryzen 7 6800H running at a clock speed of 3.2-4.2GHz. 
In the experiments, efforts are made to maintain a consistent ambient temperature to prevent CPU throttling from affecting computational speed. 
Other parameters are listed in \TableRef{sim-param}.

\input{misc/table01.tex}

\subsection{simulation of ramp merging}\label{sec-5-1}

\begin{figure*}[htbp]
\centering 
\includegraphics[scale=1.25]{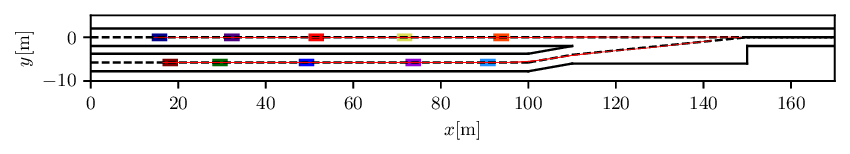}
\caption{Ramp merging area in simulation.}
\label{ramp_merge_area}
\end{figure*}

In the merging ramp scenario, the initial state of vehicles is shown in \FigRef{ramp_merge_area}, with the $x$-axis representing the lateral coordinate and the $y$-axis representing the longitudinal coordinate. 
Vehicles are represented by rectangles and differentiated by different colors. 
The red lines in the figure indicate the nominal trajectory of the vehicles. 
At that time, within the initial range of 100 meters on the main road and the ramp, there are 5 randomly generated vehicles each. 
The initial distance between vehicles is uniformly distributed in the range of 15 meters to 25 meters, and the initial velocity is uniformly distributed in the range of 40 km/h to 72 km/h. 
Next, vehicles first use the distributed trajectory planning process as shown in \AlgRef{algo-2} to complete the distributed trajectory planning, and then use \AlgRef{algo-3} every second to perform distributed model predictive control for each vehicle.

\begin{figure*}[htbp]
\captionsetup[subfloat]{captionskip=-0pt,oneside,margin={0.7cm,0cm}}
\centering
	\subfloat[\label{position-ramp-a}]{\includegraphics[width = 0.51\textwidth]{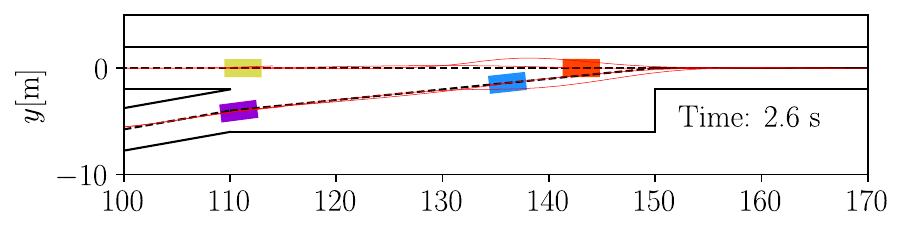}}
	\subfloat[\label{position-ramp-b}]{\includegraphics[width = 0.49\textwidth]{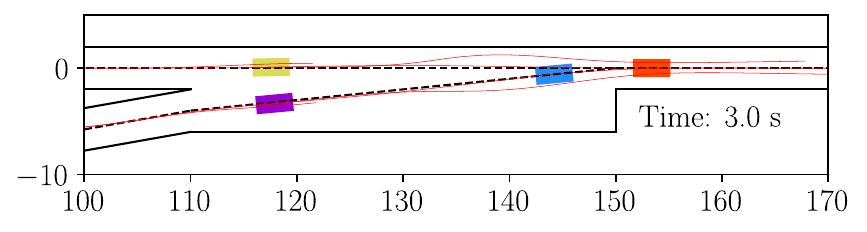}}
   \vspace{-10pt}
	\subfloat[\label{position-ramp-c}]{\includegraphics[width = 0.51\textwidth]{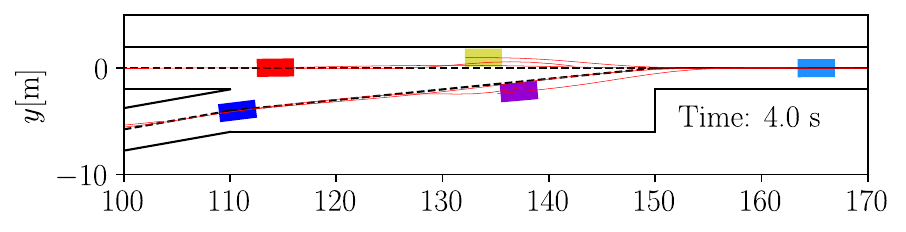}} 
	\subfloat[\label{position-ramp-d}]{\includegraphics[width = 0.49\textwidth]{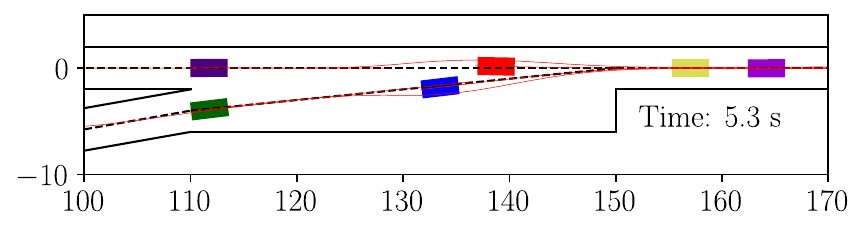}}
   \vspace{-10pt}
	\subfloat[\label{position-ramp-e}]{\includegraphics[width = 0.51\textwidth]{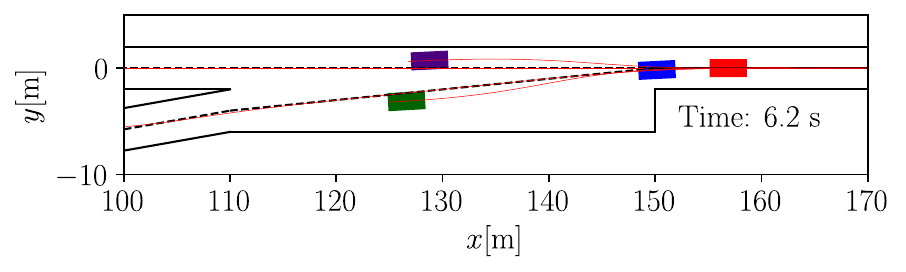}}
	\subfloat[\label{position-ramp-f}]{\includegraphics[width = 0.49\textwidth]{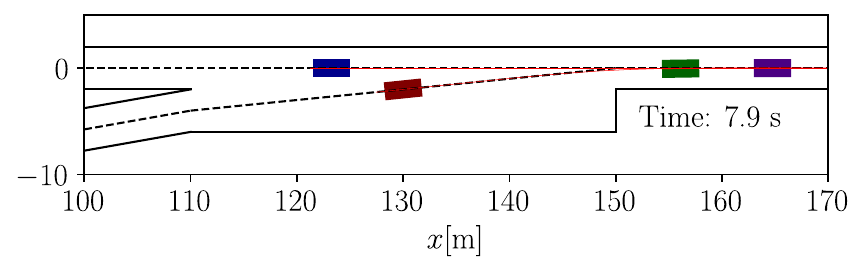}} 
\caption{The positions of vehicles at different times in the ramp merging area.}
\label{position-ramp}
\end{figure*}

\FigRef{position-ramp} shows screenshots of the status of vehicles near the acceleration lane at times 2.6, 3.0, 4.0, 5.3, 6.2, and 7.9 seconds, with the screenshot time indicated in the bottom right corner of each subplot. 
At $t=2.6$ seconds, i.e., \FigRef{position-ramp-a}, two vehicles from the main road and the ramp are gradually approaching the end of the acceleration lane. 
Since there is a safe distance between the vehicles at this time, they are essentially following the reference trajectory. 
At $t=3.0$ seconds, i.e., \FigRef{position-ramp-b}, the two vehicles from the previous image are gradually leaving the ramp, completing a safe merge. 
Meanwhile, it can be observed from the graph that the following two vehicles are entering the acceleration lane, and their nominal trajectories at this point exhibit some deviation of around 140 meters from the reference trajectory. 
This is because the constraint on $t_{M}^{i}$ established in equation \EquRef{equ-5} only ensures that these two vehicles can maintain a safe distance after point $Q$, so the safety constraint in equation \EquRef{equ-61} causes their nominal trajectories to deviate from the reference trajectory to ensure safety. 
At $t=4.0$, i.e., \FigRef{position-ramp-c}, the two vehicles from the previous subplot are traveling along the reference trajectory and maintaining a certain distance, while it can also be observed that the following two vehicles are gradually entering the acceleration lane. 
At $t=5.3$, i.e., \FigRef{position-ramp-d}, a situation similar to that at $t=4.0$ seconds occurs again. 
It can be seen from the graph that the nominal trajectories of the two vehicles near 110 meters are approximately 135 meters apart at maximum. 
However, in the actual driving process, the maximum lateral distance between the two vehicles occurs around 130 meters, corresponding to the subplot at $t=6.2$ seconds, i.e., \FigRef{position-ramp-e}. 
This is because in equation \EquRef{equ-51}, the elements in the lower right submatrix of ${{\bar{Q}}_{X}}$ are larger than the other elements on the diagonal of ${{\bar{Q}}_{X}}$, causing the nominal trajectory obtained by MPC to change with time. 
At $t=7.9$, i.e., \FigRef{position-ramp-e}, the last two vehicles in the entire scenario are gradually exiting the acceleration lane, completing a safe merge.

\begin{figure}[htbp]
\captionsetup[subfloat]{captionskip=-5pt,oneside,margin={0.8cm,0cm}}
\centering
	\subfloat[reference trajectories\label{two-traj-ref}]{\includegraphics[width = 0.50\textwidth]{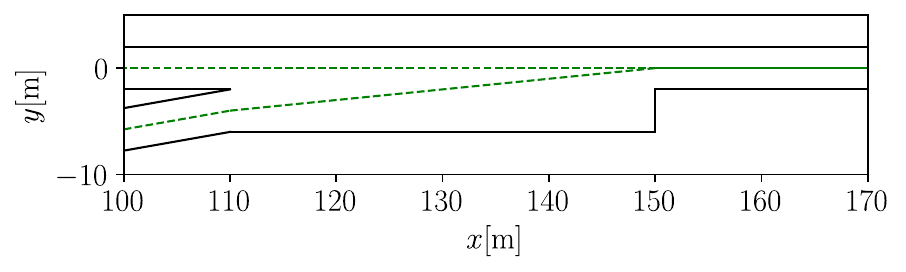}} 
   \vspace{-10pt}
	\subfloat[driving trajectories\label{two-traj-driv}]{\includegraphics[width = 0.50\textwidth]{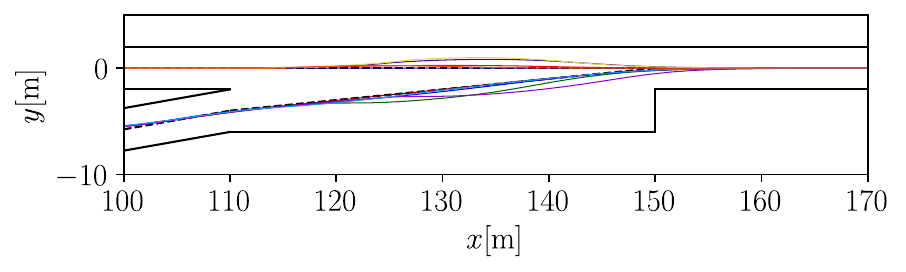}}
\caption{Reference trajectory (a) and driving trajectory (b) of vehicles in ramp merging area.}
\label{two-traj}
\end{figure}

\FigRef{two-traj} displays the reference trajectory (\FigRef{two-traj-ref}) and the actual driving trajectory of vehicles (\FigRef{two-traj-driv}). 
The green trajectory in the upper image represents the two-dimensional reference trajectory generated according to the solution of \AlgRef{algo-2} using the method outlined in \SecRef{section-2-1}. 
By comparing the two images, it can be observed that vehicles from both the main road and the ramp do not strictly follow the reference trajectory. 
The trajectory of vehicles from the main road exhibits a slight upward deviation along the reference trajectory, while the trajectory of vehicles from the ramp shows a slight downward deviation along the reference trajectory. 
This deviation occurs because the safety constraints in \EquRef{equ-61} are activated when vehicles reach around 130 meters, leading to adjustments in acceleration and steering angles to ensure safety between vehicles, as illustrated in the figure.

\begin{figure*}[hbtp]
\captionsetup[subfloat]{captionskip=-0pt,oneside,margin={0.8cm,0cm}}
\centering
	\subfloat[\label{ref_traj_iter-a}]{\includegraphics[width = 0.4\textwidth]{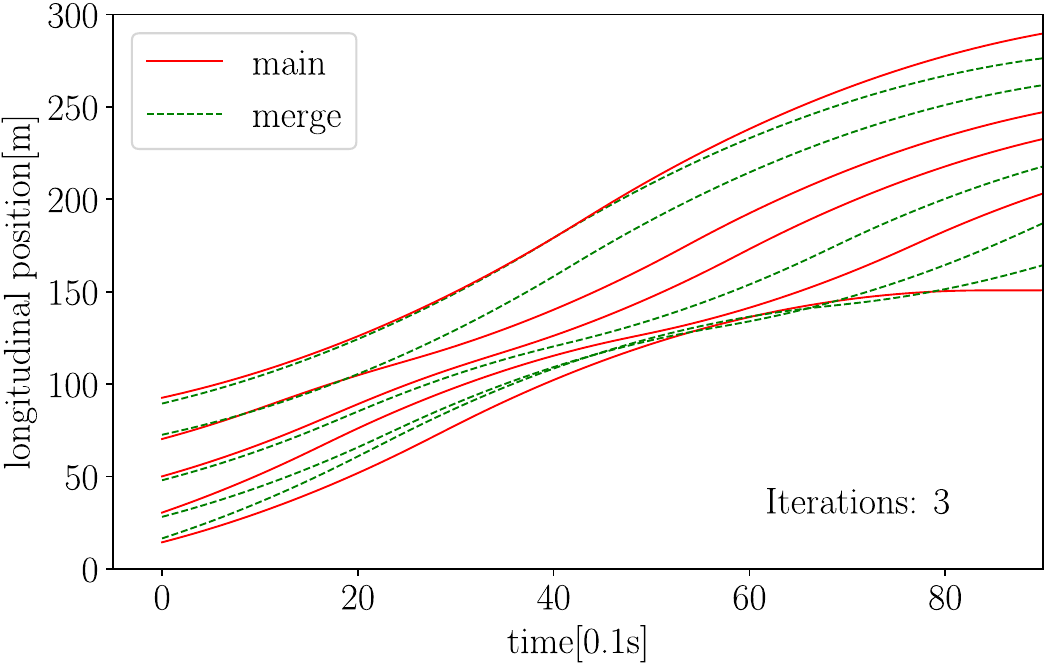}}
   \hspace{0.3cm} % 
	\subfloat[\label{ref_traj_iter-b}]{\includegraphics[width = 0.4\textwidth]{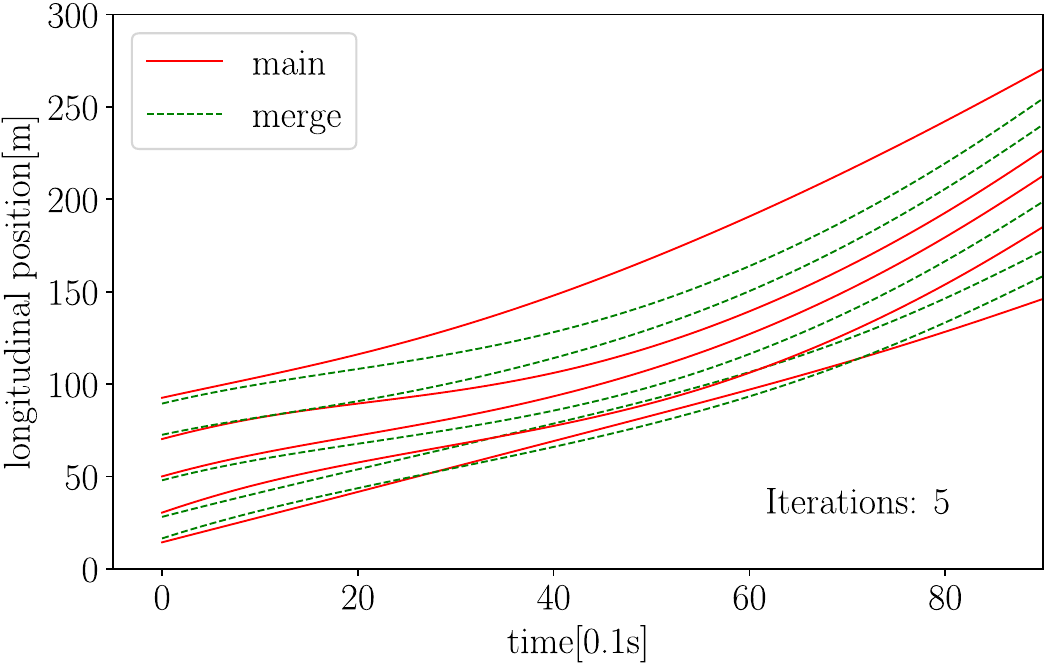}}
   \vspace{-10pt}
	\subfloat[\label{ref_traj_iter-c}]{\includegraphics[width = 0.4\textwidth]{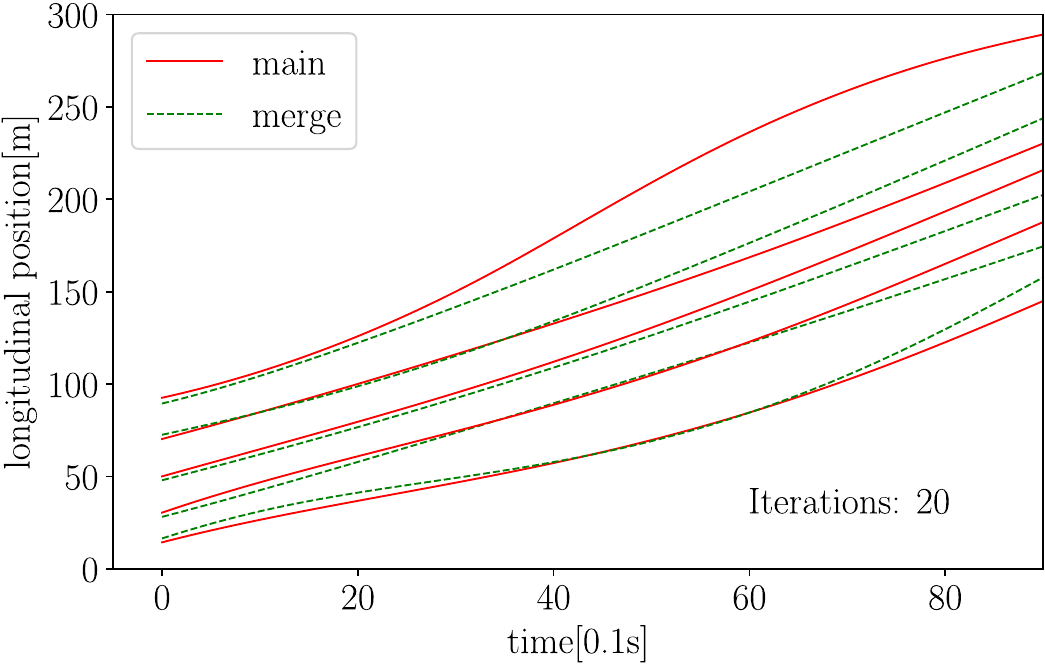}} 
   \hspace{0.3cm} % 
	\subfloat[\label{ref_traj_iter-d}]{\includegraphics[width = 0.4\textwidth]{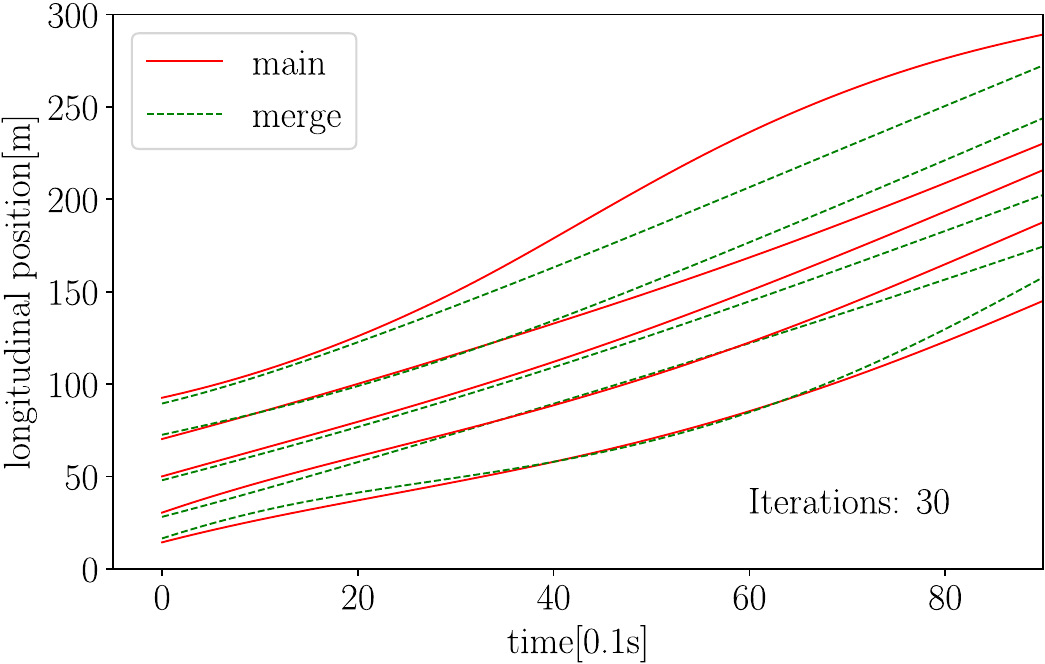}}
\caption{The evolution of the vehicle's reference trajectory with respect to the number of iterations.}
\label{ref_traj_iter}
\end{figure*}

\FigRef{ref_traj_iter} illustrates the evolution of the solution of \AlgRef{algo-2} with the change in iteration count, with the algorithm iteration count shown in the bottom right corner. 
It can be observed from the two subplots above in \FigRef{ref_traj_iter}, i.e., \FigRef{ref_traj_iter-a} and \FigRef{ref_traj_iter-b}, that the results of \AlgRef{algo-2} undergo significant changes as the iteration count increases. 
This phenomenon arises because in the initial iterations, $\sigma$ and $\rho$ are relatively small, and the "step size" of \AlgRef{algo-2} is inversely proportional to them, resulting in large differences in each iteration. 
In \cite{huang2023decentralized}, \Athr{Huang} pointed out that setting $\sigma$ and $\rho$ to smaller values in the initial iterations of \AlgRef{algo-2} can increase convergence speed. 
Therefore, the values of $\sigma$ and $\rho$ throughout the entire iteration process are shown in \TableRef{rho_sigma}. 
Additionally, it can be observed that in the initial iterations, the longitudinal distances between vehicles are relatively close, but as the iterations progress, these distances gradually increase. 
This is because during the iteration process, the longitudinal trajectory of each vehicle is essentially the solution to the optimization problem of equation \EquRef{equ-38b}. 
The solution to this problem is related to ${r}^{i,k+1}$, which in turn is related to the ${y}^{i,k+1}$ exchanged between vehicles. 
In the initial stages, vehicles do not have sufficient communication cycles, and the relevant information regarding safety constraints between vehicles has not been incorporated into ${r}^{i,k+1}$ through ${y}^{i,k+1}$. 
Therefore, the solution at this stage is close to minimizing ${{\mathcal{F}}^{i}}(U)$, i.e., without considering the safety constraints of \EquRef{equ-8}, vehicles do not maintain a safe distance between each other. 
The bottom two plots in \FigRef{ref_traj_iter}, i.e., \FigRef{ref_traj_iter-c} and \FigRef{ref_traj_iter-d}, show that the longitudinal trajectories of vehicles remain relatively unchanged between iterations 20 and 30. 
This is because as the iteration process progresses and the exchange of ${y}^{i,k+1}$ between vehicles via V2X communication occurs, \AlgRef{algo-2} gradually converges, and the distance of the longitudinal planned trajectories of vehicles gradually increases to a safe distance.

\input{misc/table02.tex}

\begin{figure}[htbp]
\centering 
\includegraphics[scale=0.5]{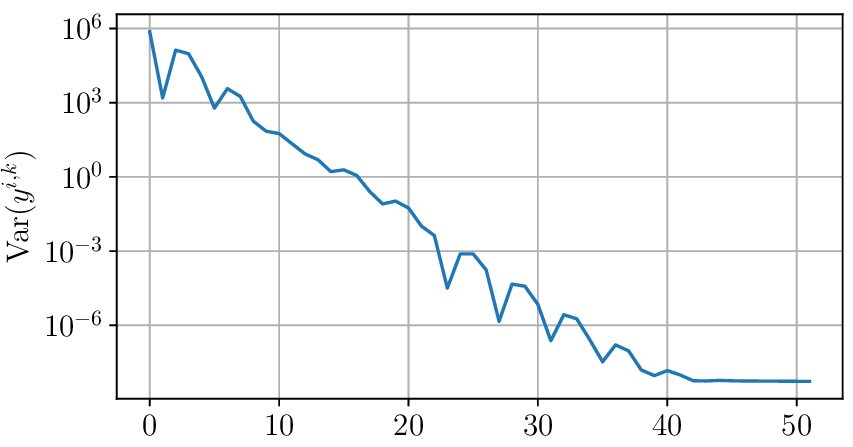}
\caption{Variance of $y^{i,k}$ with respect to the number of iterations.}
\label{var_y}
\end{figure}

\FigRef{var_y} illustrates the change in the variance of the iteration ${y}^{i,k}$ with the iteration count, where the $y$-axis is in logarithmic scale. 
The formula for calculating the variance of ${y}^{i,k}$ is given by
\begin{equation}
   Var({{y}^{i,k}})=\sum_{i=\mathcal{N}}{{{\left\| {{y}^{i,k}}-{{{\bar{y}}}^{i,k}} \right\|}^{2}}},
\end{equation}
where ${{\bar{y}}^{i,k}}=\sum\nolimits_{i=\mathcal{N}}{{{y}^{i,k}}}/\left| \mathcal{N} \right|$ represents the mean vector of ${y}^{i,k}$. 
It can be observed from \FigRef{var_y} that, in the first 20 iterations, the variance of ${y}^{i,k}$ rapidly decreases from ${10^{6}}$ to around 0.1. 
This is because in the initial iterations, $\sigma$ and $\rho$ are relatively small, leading to a larger "step size" and the absence of safety constraints among vehicles, resulting in the rapid convergence of ${y}^{i,k}$ within a certain range. 
At iteration 30, the variance of ${y}^{i,k}$ converges to around ${10^{-6}}$, and after iteration 40, the variance remains almost unchanged. 
This indicates that as the iterations progress, ${y}^{i,k}$ among vehicles become essentially equal, forming a "consensus". 
Considering real-time constraints, we choose the iteration count to be 33.

\input{misc/table03.tex}

\TableRef{average-total-time} illustrates the variation of average per-vehicle computation time and total computation time with the number of vehicles. 
$T_{avg}$ represents the time each vehicle spends on calculations per iteration, as described in \AlgRef{algo-2}. 
$T_{total}$ is the cumulative time spent by each vehicle over 33 iterations in \AlgRef{algo-2}.
It can be observed from the table that both $T_{avg}$ and $T_{total}$ increase with the number of vehicles. 
This is because as the number of vehicles increases, the dimensions of the vectors ${{p}^{i,k+1}}$, ${{r}^{i,k+1}}$, ${{s}^{i,k+1}}$, and ${{z}^{i,k+1}}$ in \AlgRef{algo-2} gradually increase, leading to increased computational workload. 
However, the increase in computation time appears relatively gradual with the increase in the number of vehicles. 
Even with 10 vehicles, the total per-vehicle computation time remains around a short 100 ms.

\subsection{simulation of DCIMPC}\label{sec-5-2}

This section compares five additional algorithms, whose names correspond to the labels in the figure below:

\begin{itemize}
   \item IPOPT: Interior Point OPTimizer (IPOPT) is a commonly used software library for large-scale nonlinear optimization of continuous systems. 
   It directly solves optimization problems with nonlinear constraints, such as \EquRef{equ-14}. 
   For vehicle dynamics constraints, if direct elimination of these constraints by function composition as
   \begin{equation}
   \begin{aligned}
   & X_{(k+1,k)}^{i}={{f}_{S}}(X_{(k,k)}^{i},U_{(k,k)}^{i}) , \\
   & X_{(k+2,k)}^{i}={{f}_{S}}({{f}_{S}}(X_{(k,k)}^{i},U_{(k,k)}^{i}),U_{(k+1,k)}^{i}) , \\
   & \vdots
   \end{aligned}
   \end{equation}
   it will makes the solution extremely slow, exceeding 1 second.
   Therefore, this section treats state variables as optimization variables and establishes constraints at each time point.

   \item LD-IPOPT: One of the main contributions of our work is the handling of nonlinear, non-convex safety constraints in \EquRef{equ-14c}. 
   To specifically compare this, the vehicle dynamics constraints are linearized and inputted together with the original nonlinear safety constraints into IPOPT for solution, termed as LD-IPOPT.

   \item SQP: Sequential Quadratic Programming (SQP) is also an algorithm used for solving non-convex optimization problems, similar to quasi-Newton methods. 
   Similar to LD-IPOPT, the vehicle dynamics constraints are linearized and inputted for solving.

   \item OSQP-CS: As mentioned in \AlgRef{algo-3} in \SecRef{sec-4-2}, the nominal trajectory serves as the starting point for OSQP warm start. 
   Thus, OSQP cold start is used as another comparative algorithm to evaluate the impact of this operation.
\end{itemize}

\begin{figure}[htbp]
\centering 
\includegraphics[scale=0.5]{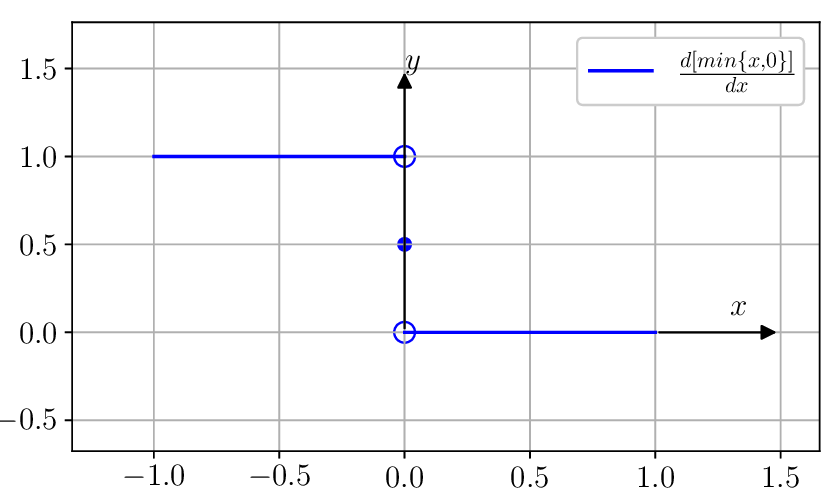}
\caption{The derivative of the function $f=\min\{0,x\}$ with respect to $x$.}
\label{derivative-fmin}
\end{figure}

Additionally, three details need to be emphasized:
\begin{itemize}
   \item Since interior point methods, i.e., as in IPOPT and LD-IPOPT, require a feasible initial solution, it is difficult to find one if safety constraints are left in the constraint set. 
   Therefore, when using these two interior point-based comparative algorithms, the unprocessed non-convex safety constraints are added to the objective function. 
   In this case, only a feasible initial solution satisfying the system constraints is needed as the starting point for interior point methods. 
   Given that the extension of the nominal trajectory in \FigRef{DCIMPC} comes from the system function $f_S$, it naturally serves as a feasible solution. 
   Therefore, the initial points in IPOPT and LD-IPOPT both adopt the nominal trajectory.

   \item The function $\tilde{D}$ in \EquRef{equ-52} is nonsmooth, and its gradient (or Jacobian matrix) is actually the subgradient. 
   The symbolic computation toolbox CasADi is used in our work to differentiate the nonsmooth function $f(x) = \min \{x, 0\}$. In the \FigRef{derivative-fmin}, the derivative of $f(x)$ at $x=0$ in CasADi is 0.5.

   \item When the computational time of the proposed algorithm at each sampled control point is analyzed in the subsequent simulation section, it includes all the time spent on each point's ${T}_{tier}$ iterations. 
   Each iteration mainly consists of two parts: first, the time taken to compute various matrices in the quadratic programming problem, such as the time for convex reconstruction of the vehicle kinematic model and safety constraints, and second, the time taken for solving with OSQP, IPOPT, or SQP.
\end{itemize}

\begin{figure}[htbp]
\captionsetup[subfloat]{captionskip=-1pt,oneside,margin={0.8cm,0cm}}
\centering
	\subfloat[T-junctions\label{T_map}]{\includegraphics[width = 0.35\textwidth]{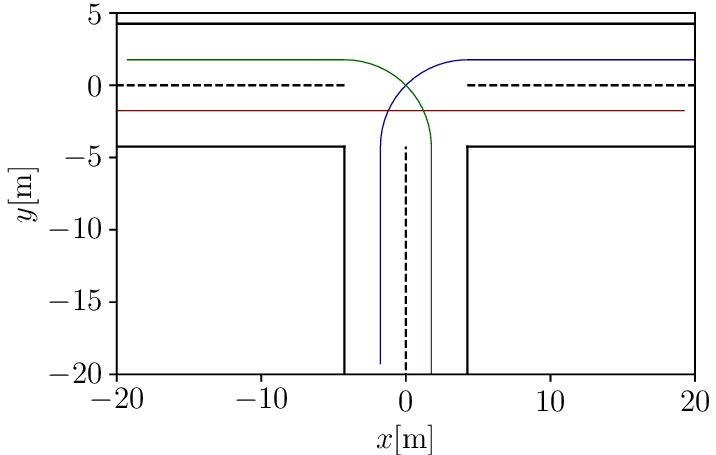}} 
   \vspace{-3pt}
	\subfloat[Crossroads\label{Crossroads-map}]{\includegraphics[width = 0.3\textwidth]{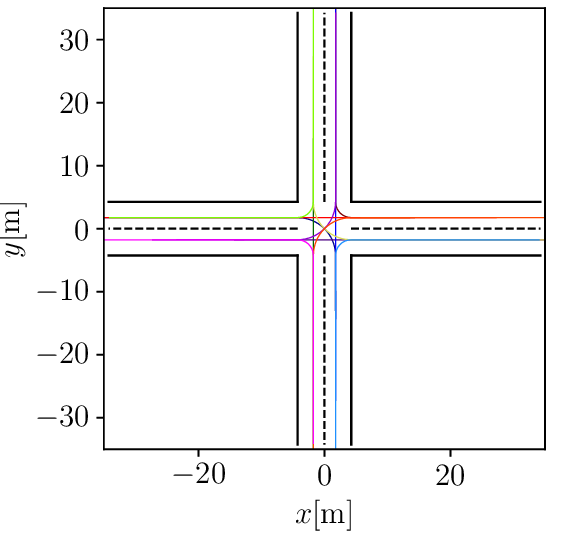}}
\caption{Reference trajectories for vehicles at intersections and T-junctions.}
\label{T_cross_map}
\end{figure}

In this subsection, we consider two scenarios as depicted in \FigRef{T_cross_map}. 
The solid black lines represent road edges or lane dividers, while the solid colored lines represent the reference trajectories of vehicles. 
The first scenario illustrates a T-junction, where vehicles from three roads intersect at the junction simultaneously. 
The second scenario is relatively more complex. 
Initially, up to three vehicles are generated in each right-hand lane of the intersection, totaling 12 vehicles, and they proceed randomly onto the other three roads. 
It can be observed from the figure that the reference trajectories intersect at the intersection of the crossroads.

\begin{figure*}[b]
\centering
   \captionsetup[subfloat]{captionskip=-0pt,oneside,margin={0.6cm,0cm}}
	\subfloat[\label{T-junctions-driving-a}]{\includegraphics[width = 0.27\textwidth]{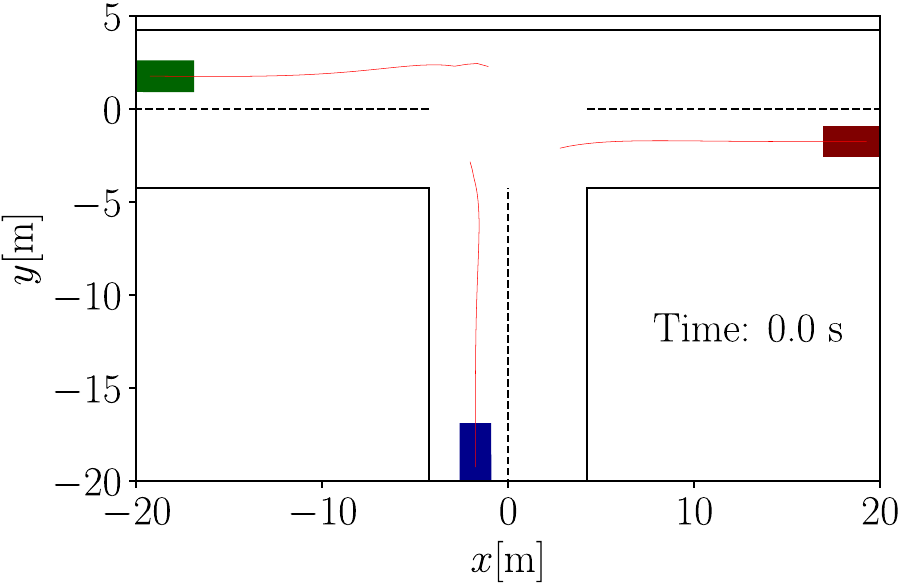}}
   \hspace{-0.1cm} % 
	\subfloat[\label{T-junctions-driving-b}]{\includegraphics[width = 0.24\textwidth]{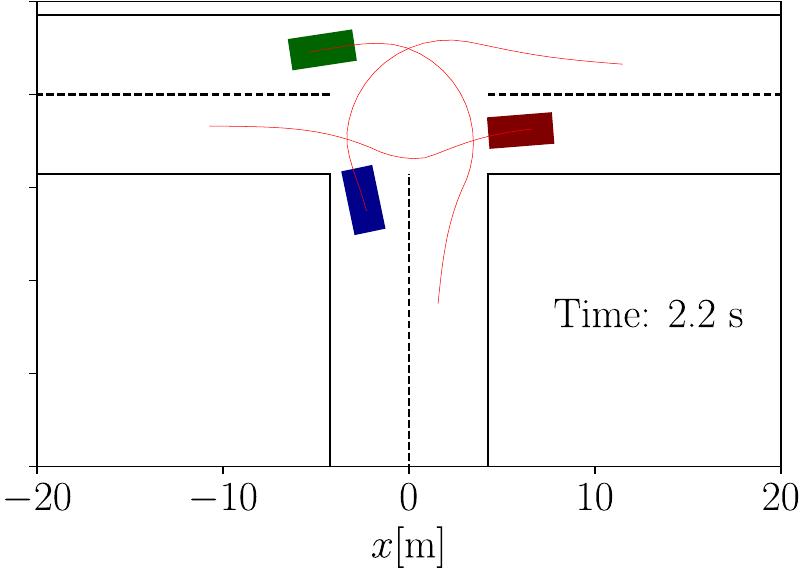}}
   \hspace{-0.1cm} % 
	\subfloat[\label{T-junctions-driving-c}]{\includegraphics[width = 0.24\textwidth]{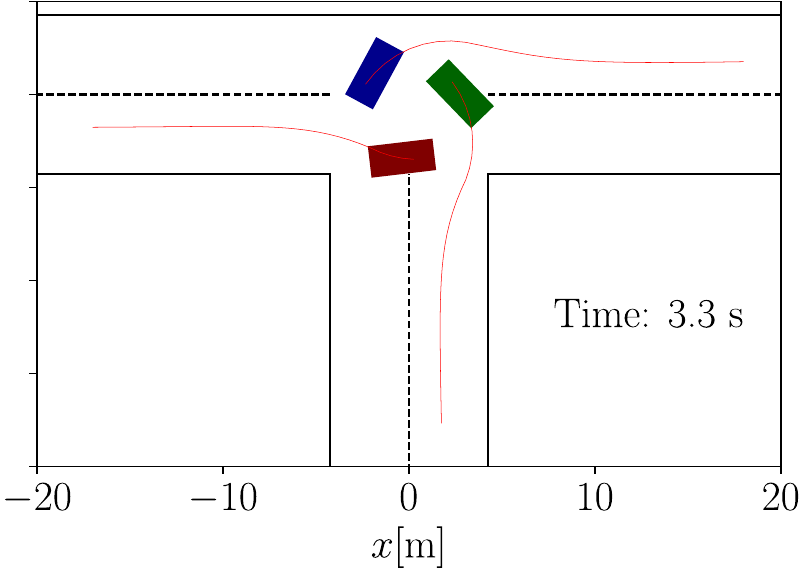}} 
   \hspace{-0.1cm} % 
	\subfloat[\label{T-junctions-driving-d}]{\includegraphics[width = 0.24\textwidth]{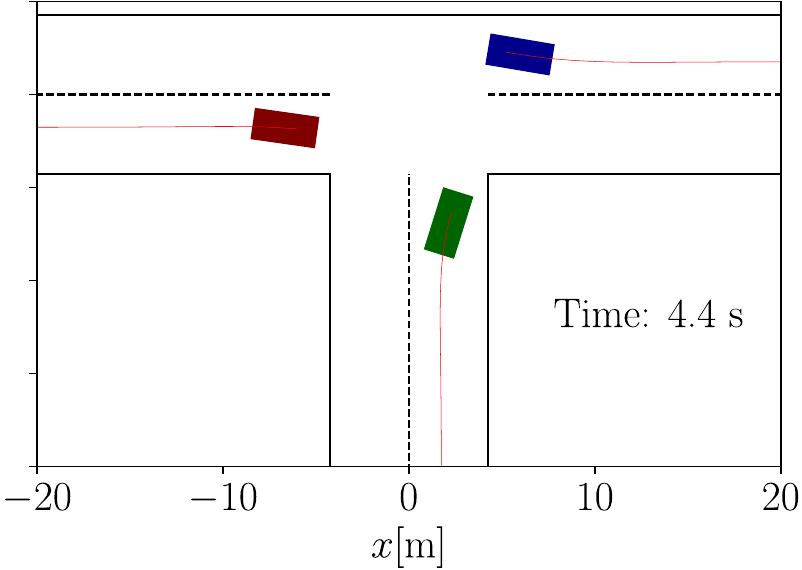}}
\caption{The evolution of the vehicle's reference trajectory with respect to the number of iterations.}
\label{T-junctions-driving}
\end{figure*}

The first scenario depicts a T-junctions. 
\FigRef{T-junctions-driving} illustrates snapshots of four representative vehicle states at times 0s, 2.2s, 3.3s, and 4.4s. 
The red lines represent the nominal trajectories of the vehicles. 
At time 0s, i.e., \FigRef{T-junctions-driving-a}, due to the constraints of the MPC horizon $T_P^2$, there is no significant avoidance behavior evident in the vehicles' nominal trajectories. 
As the vehicles moving, they gradually approach each other, entering each other's prediction range of $T_P^2$ sample points, resulting in clear avoidance behaviors in their nominal trajectories, as shown is \FigRef{T-junctions-driving-b}. 
By the time 3.3s arrives, i.e., \FigRef{T-junctions-driving-c}, all three vehicles are approaching the T-junctions under the influence of their nominal trajectories. 
However, they do not strictly adhere to the reference trajectories but rather follow trajectories closer to those predicted at 2.2s, ensuring safe travel. 
At 4.4s, i.e., \FigRef{T-junctions-driving-d}, the vehicles are gradually departing from the T-junctions. 
Since there are no additional vehicles appearing on their reference trajectories at this time, the safety constraints are not activated, indicating that their travel paths will gradually revert to following the reference trajectories.

\begin{figure}[htbp]
\captionsetup[subfloat]{captionskip=-0pt,oneside,margin={-1.3cm,0cm}}
\centering
	\subfloat[\label{T-junctions-time-a}]{\includegraphics[width = 0.5\textwidth]{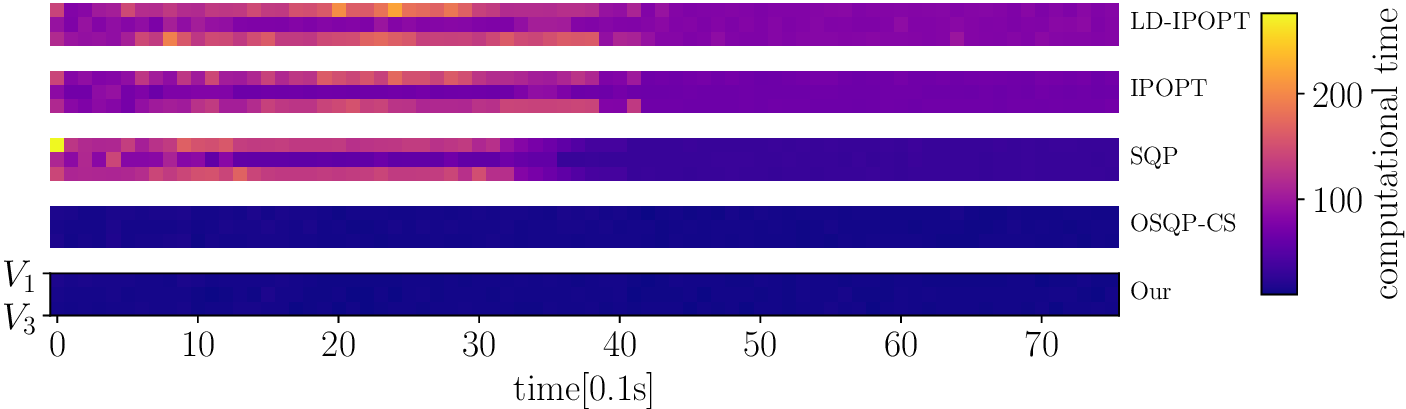}}
   \vspace{-0.1cm} % 
	\subfloat[\label{T-junctions-time-b}]{\includegraphics[width = 0.5\textwidth]{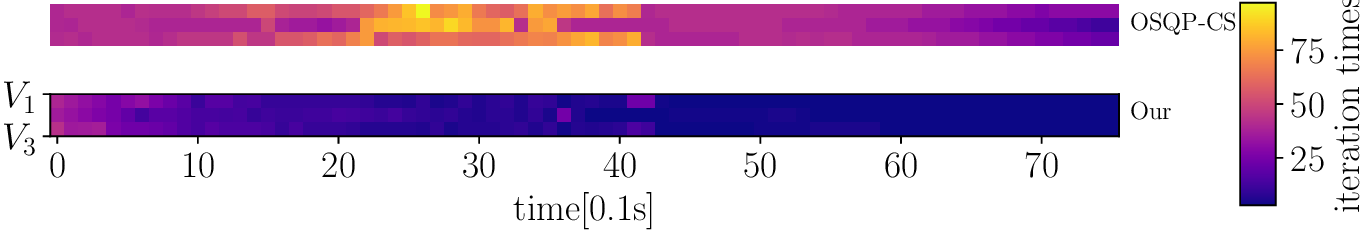}}
\caption{computational time (a) and iterations times (b) of different algorithm in T-junctions}
\label{T-junctions-time}
\end{figure}

\begin{figure}
   \centering
   \includegraphics[width=0.4\textwidth]{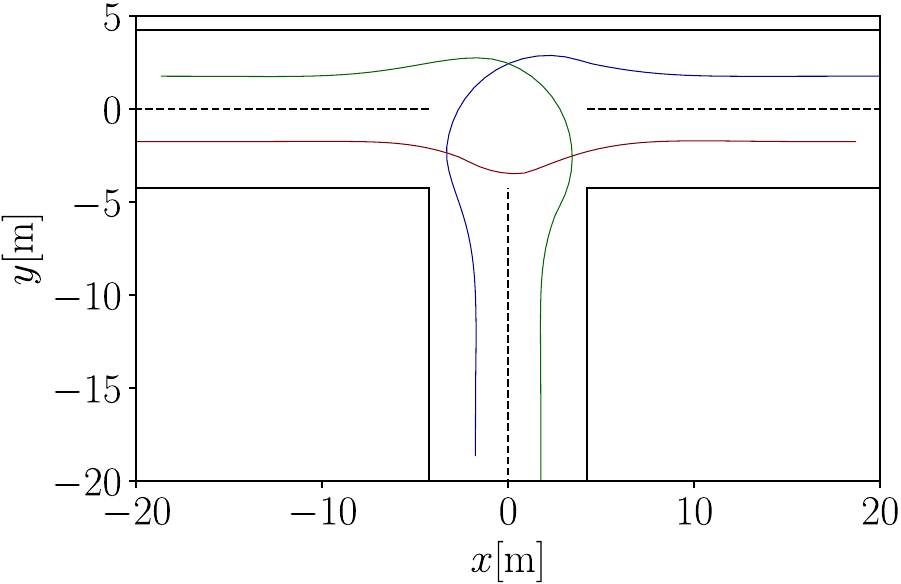}
   \caption{driving trajectories in T-junctions}
   \label{T-junctions-driving-traj}
\end{figure}

\begin{figure}[htbp]
\centering
\hspace{-1cm}
   \includegraphics[width=0.5\textwidth]{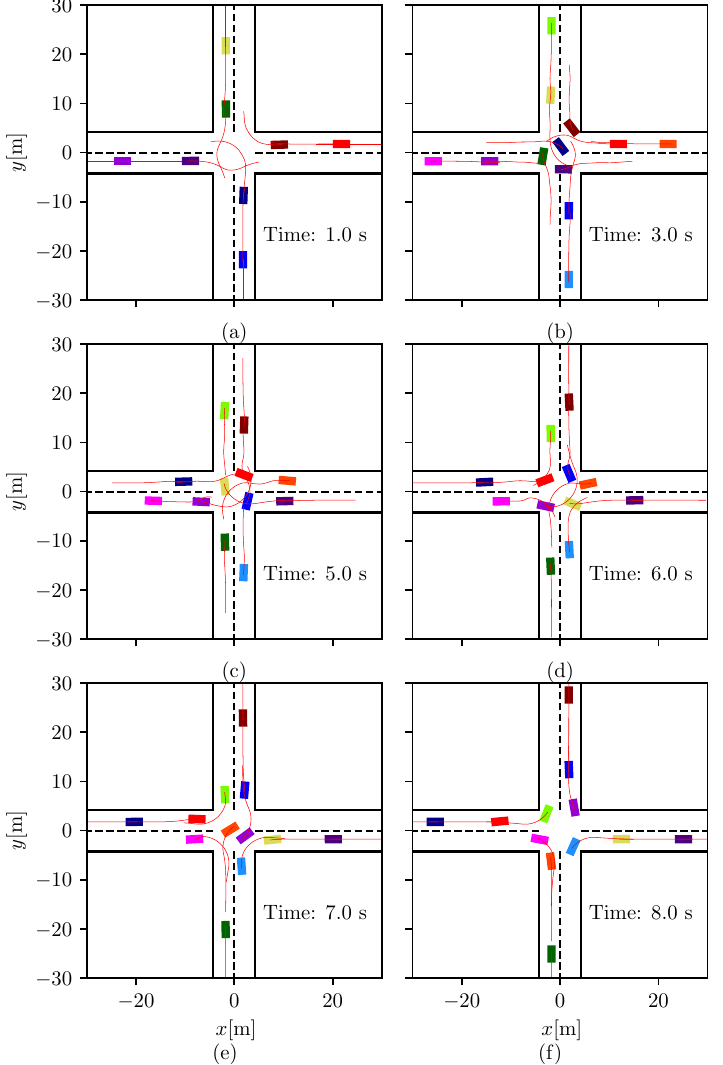}
\caption{The positions of vehicles at different times in crossroads.}
\label{cross-map}
\end{figure}

\begin{figure}
   \centering
   \includegraphics[width=0.48\textwidth]{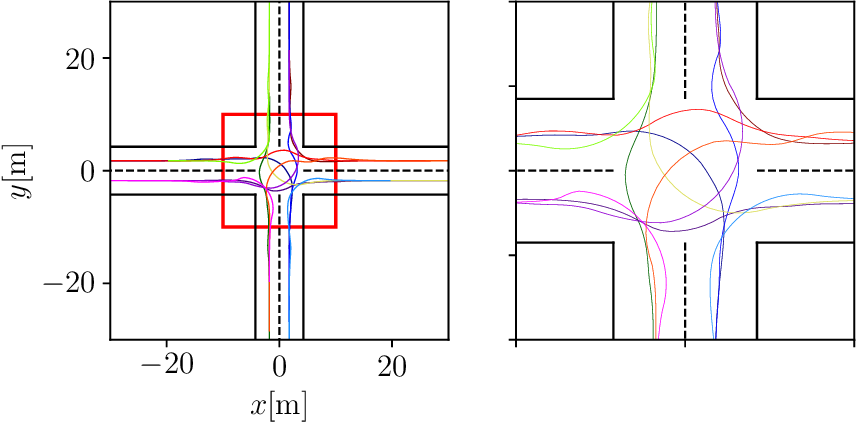}
   \caption{driving trajectories in crossroads}
   \label{crossroad-driving-traj}
\end{figure}

\begin{figure*}[hbtp]
\centering
\captionsetup[subfloat]{captionskip=-0pt,oneside,margin={1.0cm,0cm}}
	\subfloat[\label{acc-crossroad}]{\includegraphics[width = 0.4\textwidth]{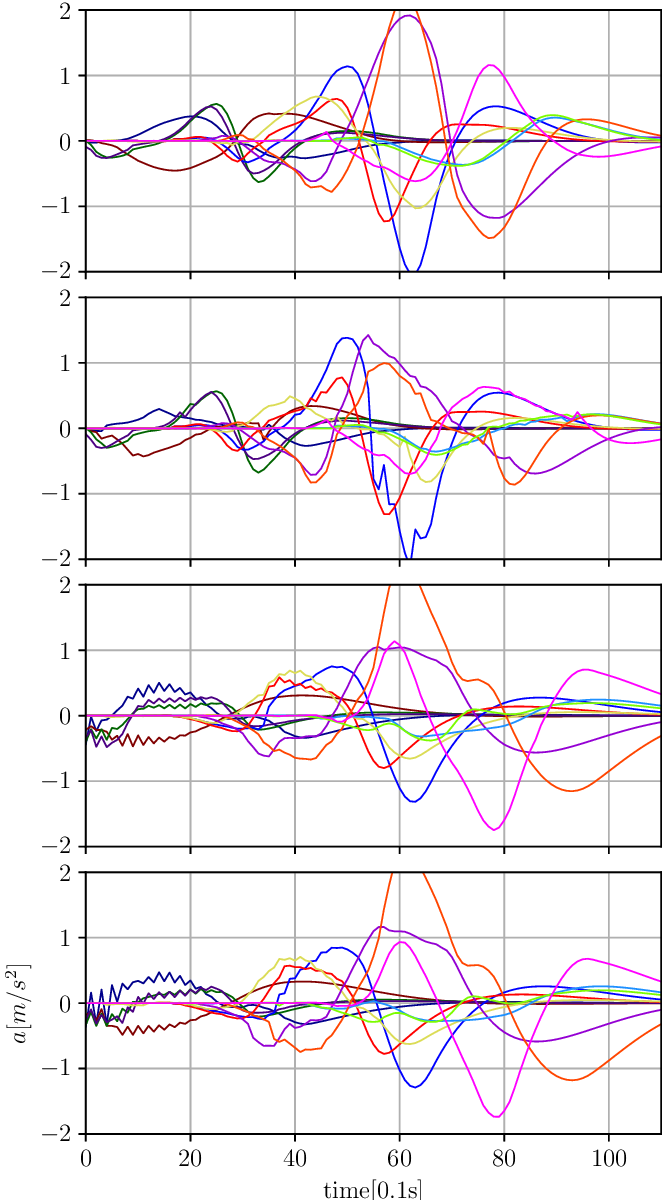}}
   \hspace{5pt}
	\subfloat[\label{steer-crossroad}]{\includegraphics[width = 0.4\textwidth]{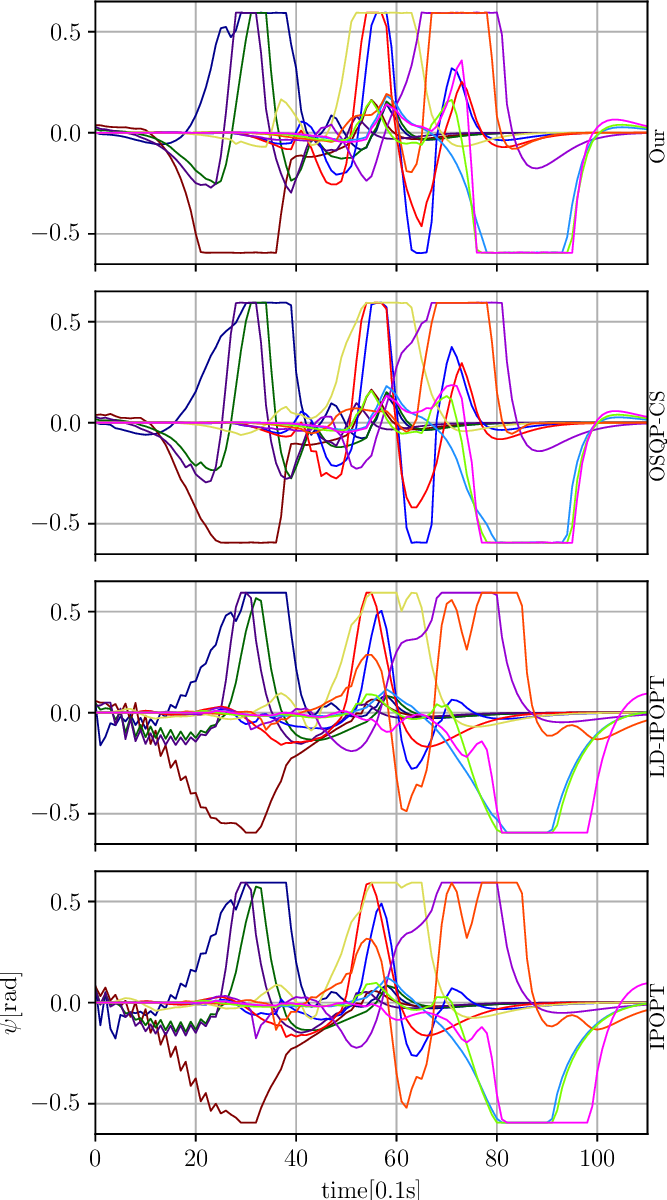}}
\caption{The positions of vehicles at different times in crossroads.}
\label{acc-steer-crossroad}
\end{figure*}

\FigRef{T-junctions-time} illustrates the algorithm's runtime (\FigRef{T-junctions-time-a}) and iteration count (\FigRef{T-junctions-time-b}) at different times for vehicles at a T-junctions. 
Each subplot corresponds to a specific algorithm, with each row representing a vehicle and time increasing from left to right. 
The color of each small rectangle represents the computation time. 
From \FigRef{T-junctions-time-a}, it can be observed that the algorithm proposed in this paper and OSQP-CS have significant advantages over the other three algorithms in terms of solving time. 
This is attributed to the handling of safety constraints by these two algorithms, which transforms problem \EquRef{equ-43} into a convex quadratic programming problem, i.e., \EquRef{equ-61}, leading to faster solutions. 
This also demonstrates that the time cost of linearizing safety constraints is much lower than the computational cost of solving non-convex problems using IPOPT and SQP. 
Additionally, considering the sampling frequency of model predictive control used in this chapter is 100ms, the solution time must be less than 100ms for practical implementation of this control scheme. 
However, as shown in \FigRef{T-junctions-time-a}, the solution times for IPOPT, linear IPOPT, and SQP algorithms mostly range from 150 to 250ms, making actual deployment challenging. 
In real-world scenarios, it may be necessary to lower the sampling frequency and utilize dedicated computing chips for acceleration. 
\FigRef{T-junctions-time-a} also indicates that instances of longer solution times mostly occur between 0 and 4 seconds, as vehicles gradually approach each other and reference trajectories intersect. 
As a result, safety constraints are activated, requiring more iterations for the algorithm to converge.

Due to the small difference in solving time between the algorithm proposed in this paper and OSQP-CS, in order to highlight the benefits of using nominal trajectory warm-starting for OSQP, the iteration counts of these two algorithms for OSQP are shown in \FigRef{T-junctions-time-b}. 
It can be observed from the graph that the iteration count of the algorithm proposed in this paper is significantly lower than that of OSQP-CS. 
This is because the warm-starting point of the algorithm in this paper is closer to the optimal point of \EquRef{equ-61}, thus requiring fewer iterations to meet the convergence criteria. 
Additionally, the graph also indicates that the algorithm proposed in this paper only has higher iteration counts at the beginning, which is due to the lack of a good warm-starting point at the initial stage before solving. 
Furthermore, from the graph, it can be seen that the iteration counts for OSQP-CS are higher between 2 to 4s, which aligns with the analysis in the preceding text, indicating that the activation of safety constraints makes solving problem in \EquRef{equ-61} more complex, requiring more iterations. 
Moreover, integrating the information from both graphs in \FigRef{T-junctions-time}, it can be observed that the computation time of the algorithm proposed in this paper is predominantly occupied by the time spent on linearization and convex reconstruction, which accounts for a significant portion of the total computation time. 
However, this transformation converts problem \EquRef{equ-43} into a convex optimization problem, enabling the use of a more efficient solver, OSQP, ultimately reducing the total computation time.

\FigRef{T-junctions-driving-traj} illustrates the actual trajectories of vehicles at a T-junction. 
As depicted in the figure, vehicles initially deviate to the left and then to the right along their respective reference paths, resulting in the curved trajectory shown in the right panel of \FigRef{T-junctions-driving-traj}. 
This deviation along the reference paths occurs because the exchange of nominal trajectories among vehicles spreads the deviation caused by safety constraints along the reference paths among the three vehicles. 
Consequently, each vehicle only needs to make minor adjustments to avoid collision, leading to the observed phenomenon.

\FigRef{cross-map} displays screenshots of the driving situation of 12 vehicles at a crossroads scenario. 
The specific time is shown in the down right corner, and the red lines represent nominal trajectories. 
At 1s, i.e., \SubFigRef{cross-map}{a}, the nominal trajectories of the first four vehicles from each of the four roads have already mutually yielded under the advance planning of model-predictive control, forming trajectories that allow safe passage. 
It can be observed that the nominal trajectory of vehicles from the right-side road is closer to the road edge than the reference trajectory. 
This alleviates pressure on the trajectory solving for the first vehicle from the other three roads, as this vehicle can receive nominal trajectories from the other three vehicles through vehicle-to-vehicle communication, achieving multi-vehicle cooperation. 
By the time 3s are reached, the first batch of vehicles has essentially crossed the crossroads, as shown in \SubFigRef{cross-map}{b}. 

Then, to increase complexity, connections are established between the second and third batches of vehicles by adjusting the reference trajectories. 
From the two subplots in \SubFigRef{cross-map}{c} and \SubFigRef{cross-map}{f}, it can be seen that the second vehicle from the left-side road will eventually reach the position of the third vehicle from the top road under the influence of the reference trajectory. 
Similarly, the third vehicle from the right-side road will eventually reach the position of the second vehicle from the bottom road. 
The two images at 5s (\SubFigRef{cross-map}{c}) and 6s (\SubFigRef{cross-map}{d}) reflect two relatively complex time points during the entire driving process, during which vehicles maintain safe distances between each other and from the road edges. 
These are reflected in the safety constraint and reference trajectory parts of problem in \EquRef{equ-61}. 
By adjusting the weights of these two components, adaptation to different scenarios is possible. 
As shown in \SubFigRef{cross-map}{f}, by 8 seconds, all 12 vehicles have safely crossed the crossroads, demonstrating the effectiveness of the algorithm proposed in this paper even in relatively complex scenarios.

\FigRef{crossroad-driving-traj} illustrates the actual driving trajectories of vehicles controlled using our algorithm. 
It should be noted that the SQP algorithm failed to converge in this scenario, so the experiments concerning the crossroad do not include results from the SQP algorithm. 
In \FigRef{crossroad-driving-traj}, it can be observed that vehicles deviate from the reference trajectory to yield to each other by continuously adjusting the rate of change of heading angle to alter their heading angles. 
The square region in the center of the crossroad, marked by the red box in \FigRef{crossroad-driving-traj}, is enlarged. 
By comparing with the reference trajectories in \FigRef{T_cross_map}, it can be seen that vehicles turning right tend to hug the road edge more closely than the reference trajectory to avoid other vehicles in the center, while vehicles turning right or going straight form curved driving trajectories to ensure safe passage.

\FigRef{acc-steer-crossroad} illustrates the variations of acceleration (\FigRef{acc-crossroad}) and steering angle (\FigRef{steer-crossroad}) over time for vehicles controlled by different algorithms in a intersection scenario. 
Regarding the acceleration curve in \FigRef{acc-crossroad}, vehicles controlled by the our algorithm exhibit higher acceleration compared to those controlled by the OSQP-CS algorithm, yet they demonstrate better stability. 
This is attributed to the warm-starting approach utilized in this paper, which enables successive optimal solutions to be closer to each other, thereby achieving smoother control outputs. 
It can be observed that both the our algorithm and OSQP-CS maintain relatively stable acceleration and steering angles at the beginning stage under the influence of iterative model predictive control, compared to the other two algorithms. 
Furthermore, whether in terms of acceleration or steering angle, by the time of 10s, both our algorithm and OSQP-CS converge faster to 0 compared to the other two algorithms, indicating their ability to track reference trajectories more quickly while avoiding collisions.

\begin{figure}[htbp]
\captionsetup[subfloat]{captionskip=-0pt,oneside,margin={-1.3cm,0cm}}
\centering
	\subfloat[\label{crossroad-time-a}]{\includegraphics[width = 0.5\textwidth]{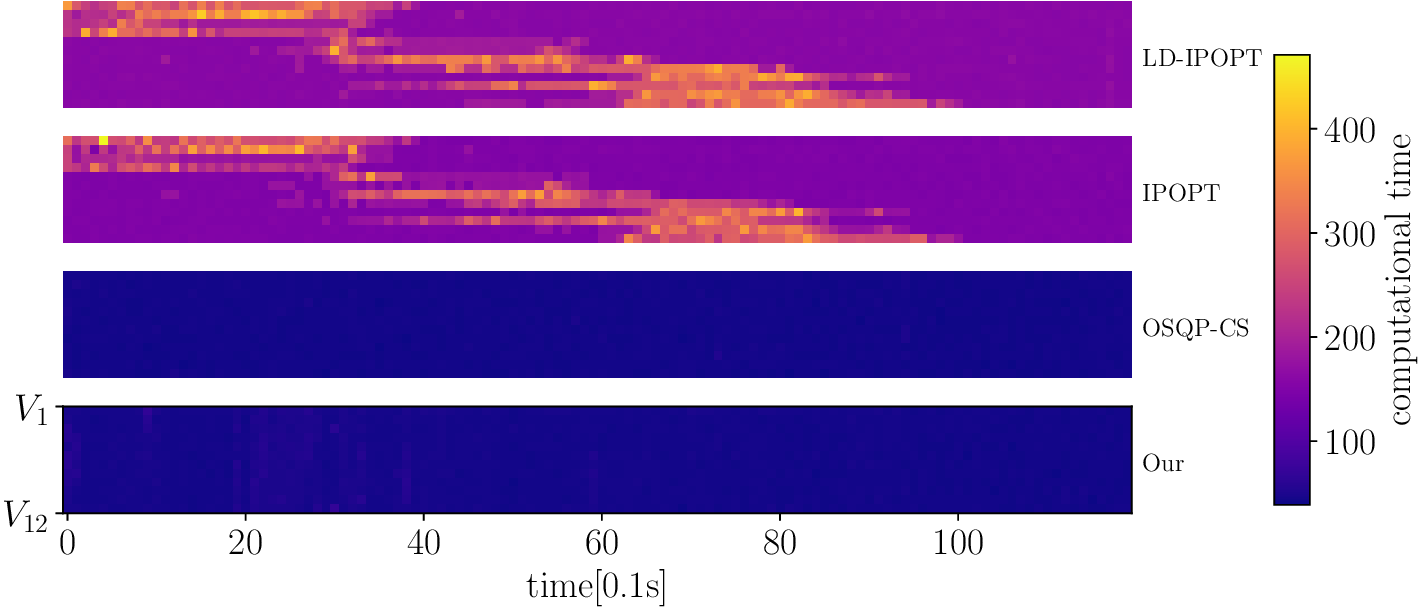}}
   \vspace{-0.1cm} % 
	\subfloat[\label{crossroad-time-b}]{\includegraphics[width = 0.5\textwidth]{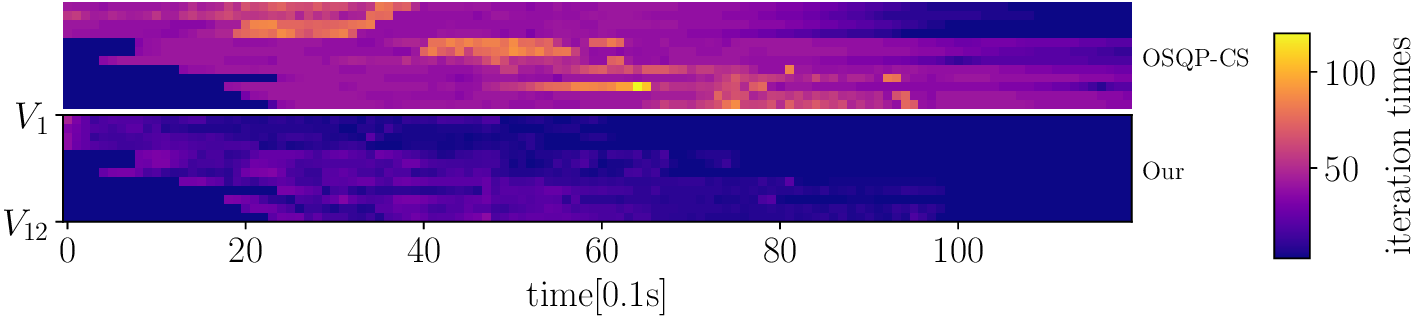}}
\caption{computational time (a) and iterations times (b) of different algorithm in T-junctions}
\label{crossroad-time}
\end{figure}

\begin{figure}[htbp]
\captionsetup[subfloat]{captionskip=-0pt,oneside,margin={0.5cm,0cm}}
\centering
	\subfloat[\label{min_dist-a}]{\includegraphics[width = 0.45\textwidth]{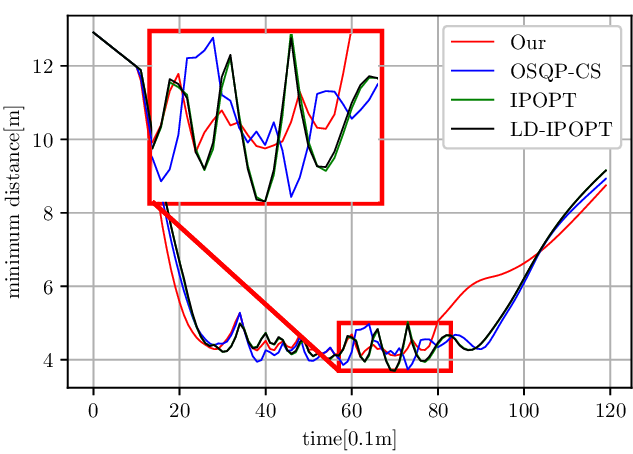}}
   \vspace{-0.2cm} % 
	\subfloat[\label{min_dist-b}]{\includegraphics[width = 0.47\textwidth]{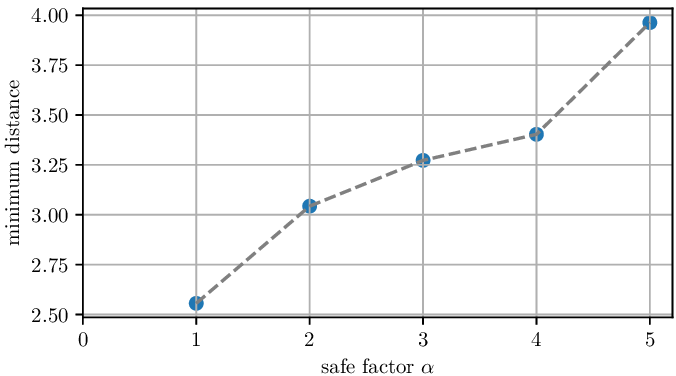}}
\caption{computational time (a) and iterations times (b) of different algorithm in T-junctions}
\label{min_dist}
\end{figure}

\FigRef{crossroad-time} depicts the computational time (\FigRef{crossroad-time-a}) and the comparison of iteration times (\FigRef{crossroad-time-b}) between our algorithm and OSQP-CS in the intersection scenario. 
The meaning of the figure is identical to that of \FigRef{T-junctions-time}. 
From \FigRef{crossroad-time-a}, it can be observed that both our algorithm and OSQP-CS maintain a significant lead in computational speed over IPOPT with interior-point method and linear-IPOPT. 
Moreover, the computational time is concentrated around 50ms, making it feasible for deployment in real-time under the sampling frequency and hardware conditions utilized by the authors. 
By examining figure of IPOPT and linear-IPOPT, it is noticeable that the longer computational times are mainly concentrated in three areas: from vehicle 1 to vehicle 4 between 0 to 4 seconds, from vehicle 5 to vehicle 8 between 4 to 6 seconds, and from vehicle 9 to vehicle 12 between 6 to 10 seconds. 
The latter two regions are interconnected, which is precisely the outcome discussed in explaining \FigRef{cross-map}, where the use of reference trajectories introduces connections among the latter two batches of vehicles. 
Since the difference in computational time between the algorithm proposed in this paper and OSQP-CS is still relatively small, \FigRef{crossroad-time-b} below supplements a comparison of iteration times. 
From this figure, it can be observed that in the intersection scenario, the reduction in iteration times due to the OSQP warm start using nominal trajectories is more pronounced compared to the T-junction scenario. 
This is because the complexity of the intersection scenario is significantly higher than that of the T-junction, requiring more iterations for the cold start to reach the optimal point. 
It can be seen from \FigRef{crossroad-time-b} that OSQP-CS reaches over 100 iterations in the worst-case scenario, while the algorithm proposed in this paper does not exceed 60 iterations. 
Additionally, it can be observed from \FigRef{crossroad-time-b} that the positions with higher OSQP iteration times vary for different vehicles but are all concentrated near the intersection. 
This is because when the length of the predicted trajectories of the vehicles does not cover the intersection, the nominal trajectories do not intersect, meaning that the safety constraints in equation \EquRef{equ-61} are not activated. 
As time progresses and the distances between vehicles shorten, the safety constraints gradually become activated. 
These old nominal trajectories distributed along the reference trajectory require more iterations as the distance from the optimal point increases when used as warm start points, thereby requiring more iterations.

\FigRef{min_dist-a} illustrates the relationship between the minimum distance between vehicles and time for different algorithms in a crossroad scenario. 
It can be observed from the graph that the minimum distances for various algorithms are roughly the same, but our algorithm is slightly smaller than the others before 3 seconds. 
This may be because our algorithm introduces relatively large errors when approximating the safety constraints. 
However, at this point, the vehicles still maintain a relatively safe distance from each other, and the advantage in computation time far outweighs that of other algorithms. 
Between 6 and 8 seconds, the minimum distance between vehicles reaches its minimum value, and at this time, the minimum distances for various algorithms are basically the same. 
However, from 8 to 10 seconds, there is a significant difference between our algorithm, OSQP-CS, and the two algorithms based on the interior point method, due to the different driving trajectories resulting from different solutions.

The safety coefficient $\alpha$ is a crucial hyperparameter that controls the weight of safety constraints and reference trajectories in the optimization objective. 
\FigRef{min_dist-b} illustrates the variation in the minimum distance between vehicles as different safety coefficients ranging from 1 to 5 are applied. 
As expected, with an increase in the safety coefficient, the minimum distance gradually increases. 
This is because as $\alpha$ increases, the proportion of the safety constraint term in \EquRef{equ-61} gradually increases, thereby causing the nominal trajectory to deviate further from the reference trajectory to increase the distance between vehicles.

\begin{figure}
   \centering
   \includegraphics[width=0.48\textwidth]{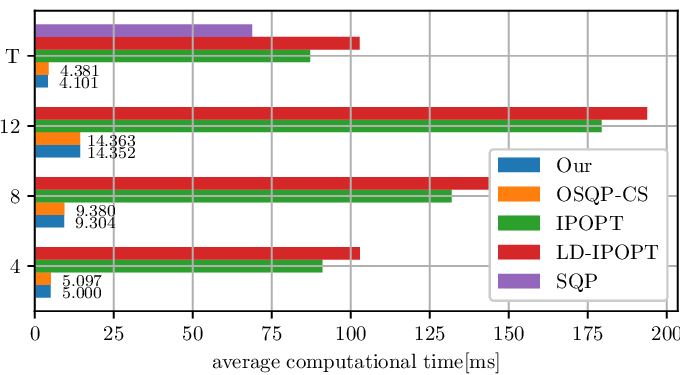}
   \caption{driving trajectories in crossroads}
   \label{avg_comp_time}
\end{figure}

\FigRef{avg_comp_time} presents a comparison of the average computation time per control step for different algorithms in various scenarios. 
The "8" and "4" represent the results excluding the last batch of vehicles and the last two batches of vehicles at the intersection, respectively. 
Since SQP failed to solve in the intersection scenario, the results including the SQP algorithm are only included in the "T" section. 
It can be observed from the graph that although the algorithm proposed in this paper has significantly fewer iterations compared to OSQP-CS, the average computation time and the total algorithm time difference between them are minimal. 
This is because the solving time of OSQP accounts for only a small fraction of the total computation time, with the majority of time spent on solving the Jacobian matrix. 
Additionally, it can be seen from the graph that the computation time is closely correlated with the number of vehicles. 
This is because as the number of vehicles increases, the complexity of the problem increases, leading to a synchronous increase in the computational effort required to solve the optimization problem, resulting in longer solving times. 
Furthermore, from the average computation time, it can be observed that the algorithm proposed in this paper can quickly solve within 25 milliseconds, thus meeting the requirements of higher sampling frequency for model predictive control. 
However, considering that communication still requires some time, this paper sets it to 100 milliseconds.

%%%%%%%%%%%%%%%%%%%%%%%%%%%%%%%%%%%%%%%%%%%%%%%%%%%%%%%%%%%%%%%%%%%%%%%%%%%%%%%%%%%%%%%%%%%%%%%%%%%%%%%%%%%%%%%%%%%%%%
\section{Conclusions}\label{sec-6}
This paper presents a multi-vehicle cooperative control framework for ramp merging areas based on V2X assistance. 
With V2X assistance, vehicles can achieve cooperative behavior in a distributed manner through information exchange. 
To solve this complex ramp merging problem, we divided it into two steps.
In the first step, considering the characteristics of the ramp, the multi-vehicle ramp merging problem is formulated as a centralized optimization problem at a macro level. 
Then, the problem is decomposed using Lagrangian duality theory and ADMM algorithm, enabling it to be solved by vehicles distributed in the scene under the assistance of vehicular networks. 
In the second step, based on the planned trajectories, distributed cooperative control is conducted. 
Vehicles communicate with each other by V2X and establish cooperative relationships by sharing nominal trajectories, and convex reconstruction of safety constraints is performed to transform the locally non-convex optimal control problem of vehicles into a quadratic programming problem for rapid solution. 
Simulation in ramp scenarios validates the effectiveness of the proposed two step approach and the convergence of the distributed algorithm. 
Additionally, a comparison of the solution time under different numbers of vehicles demonstrates the feasibility of practical deployment. 
Furthermore, experiments are conducted to enhance the efficiency of the distributed cooperative control method without compromising solution performance.

\ifCLASSOPTIONcaptionsoff
  \newpage
\fi

\bibliographystyle{template/IEEEtran}
\bibliography{template/refs}

\end{document}

%% file: template/package.tex
\usepackage{cite}
\usepackage{graphicx}
\usepackage{amsmath}
\usepackage{booktabs}

\usepackage[caption=false]{subfig}
\usepackage{stfloats}

\usepackage{hyperref}
\usepackage{arydshln}
\usepackage{cuted}
\usepackage{lettrine}
\usepackage{xcolor}
\usepackage[normalem]{ulem}
\usepackage{soul}

\usepackage{hyperref}
\hypersetup{
    colorlinks = true, 
	linkcolor = blue,
	urlcolor = blue,
    citecolor = blue
}

\usepackage{amsthm}
\newtheorem{remark}{Remark}

\usepackage[norelsize, linesnumbered, ruled, lined, boxed, commentsnumbered]{algorithm2e}      
\SetKwRepeat{Do}{do}{while}%
\SetKw{Kwin}{ in } 
\SetKw{KwTo}{ to }
\SetKw{KwGoto}{goto }
\SetKw{Init}{Initialize: }

\usepackage{setspace}

\newcommand{\EquRef}[1]{Eq. (\ref{#1})}
\newcommand{\SecRef}[1]{Sec. \ref{#1}}

\newcommand{\TableRef}[1]{Tab. \ref{#1}}
\newcommand{\FigRef}[1]{Fig. \ref{#1}}
\newcommand{\SubFigRef}[2]{Fig. \ref{#1}{\textcolor{blue}{#2}}}
\newcommand{\Athr}[1]{{#1} \emph{et al.}}
\newcommand{\AlgRef}[1]{Alg. \ref{#1}}

%% file: misc/algo01.tex
\begin{algorithm}
    \SetAlgoLined
    \KwIn{Hyper parameters $\sigma > 0$ and $\rho > 0$}

    initialization $y^{i,0}$, $p^{i,0}$ and $s^{i,0}=0$\;
    \Do{until termination criteria}{
    exchange ${{y}^{i,k}}$ between all vehicles\;
    ${{p}^{i,k+1}}\leftarrow {{p}^{i,k}}+\rho \sum_{j\in {{\mathcal{N}}_{i}}}{({{y}^{i,k}}-{{y}^{j,k}})}$\;
    ${{s}^{i,k+1}}\leftarrow {{s}^{i,k}}+\sigma ({{y}^{i,k}}-{{z}^{i,k}})$\;
    $
    {{y}^{i,k+1}} \leftarrow 
    \underset{y}{argmin}\,\left\{ \begin{aligned}
    &\frac{\sigma }{2}{\left\| y-{{z}^{i,k}} \right\|}^{2} + \left\langle y,{{p}^{i,k+1}}+{{s}^{i,k+1}} \right\rangle\\
    &\rho \sum\limits_{j\in {{\mathcal{N}}_{i}}}{{{\left\| y-\frac{{{y}^{i,k}}+{{y}^{j,k}}}{2} \right\|}^{2}}} + d_{\mathcal{L}}^{i,1}(y)
    \end{aligned} \right\}
    $\;
    $
     {{z}^{i,k+1}} \leftarrow 
     \underset{z}{argmin}\,\left\{
        d_{\mathcal{L}}^{i,2}(z)-\left\langle z,{{s}^{i,k+1}} \right\rangle
        +\frac{\sigma }{2}{{\left\| z-{{y}^{i,k+1}} \right\|}^{2}} 
    \right\}
    $\;
    }
    \caption{Consensus ADMM}
    \label{algo-1}
\end{algorithm}

%% file: misc/algo02.tex
\begin{algorithm}
    \SetAlgoLined
    \KwIn{Hyper parameters $L_1$, $L_2$, $c_1$, $c_0$, $\sigma > 0$ and $\rho > 0$}
    
    $N$ vehicles transmit their coordinate through \textbf{V2X} to confirm their merging sequence $\mathcal{S}^i$\;
    $N$ vehicles initialize $y^{i,0}$, $p^{i,0}$ and $s^{i,0}=0$\;
    \Do{\textnormal{until termination criteria}}{
    Exchange ${{y}^{i,k}}$ between all vehicles\;
    ${{p}^{i,k+1}}\leftarrow {{p}^{i,k}}+\rho \sum_{j\in \mathcal{N}_i} (y^{i,k} - y^{j,k})$\;
    ${{s}^{i,k+1}}\leftarrow {{s}^{i,k}}+\sigma ({{y}^{i,k}}-{{z}^{i,k}})$\;
    $r^{i,k+1} \leftarrow \sigma z^{i,k} + \rho \sum_{i \in \mathcal{N}} - \left( b^i + p^{i,k+1} + s^{i,k+1} \right)$\;
    Solve problem \EquRef{equ-39} to get $U^{i,k+l}$\;
    Calculate $y^{i,k+l}$ by \EquRef{equ-38a}\;
    Calculate $z^{i,k+l}$ by \EquRef{equ-42}\;
    }
    \caption{Trajectory Planning}
    \label{algo-2}
\end{algorithm}

%% file: misc/algo03.tex
\begin{algorithm}
  \SetAlgoLined
  \SetKwInOut{Input}{Input}
  \SetKwInOut{Output}{Output}
  \Input{safe factor $\alpha > 0$, iter times $T_{iter}$, predict length $T_P^2$}

  $k=0$, reference trajectory $\tilde{X}^i$ in \AlgRef{algo-2}\;
  
  \Repeat{driving termination}{
    $C = T_{iter}$\;
    \Repeat{$C=0$}{
        Transmit nominal trajectory $\bar{X}^i_k$ of time step $k$\;
        Receive nominal trajectories $\bar{X}^j_k$, where $j \in \mathcal{N}_i$\;
        Calculate matrixs $\mathcal{P}$ and $\mathcal{Q}$ in \EquRef{equ-62}\;
        Input $\bar{X}^i_k$ to OSQP for warm start\;
        Solve QP in \EquRef{equ-61} and get $U^{i*}_{(k+l,k)}$\;
        Generate $\bar{X}^i_k$ based on $U^{i*}_{(k+l,k)}$\;
        $C \leftarrow C -1$\;
    }
    Execute $U^{i*}_{(k,k)}$\;
    Generate $\bar{X}^i_{k+1}$ based on $\bar{X}^i_k$\;
    $k \leftarrow k + 1$
  }
  \caption{DCIMPC of vehicle $i$}
  \label{algo-3}
\end{algorithm}

%% file: misc/table01.tex
\begin{table*}[htbp]
\caption{Simulation Parameters}
\centering 
\renewcommand{\arraystretch}{1.1}
\begin{tabular}{ll|ll|ll|ll}
\hline
\textbf{parameters} & \textbf{values}  & \textbf{parameters} & \textbf{values}             & \textbf{parameters}      & \textbf{values}                              & \textbf{parameters}      & \textbf{values}                                           \\ \hline
$T_S$     & $0.1$s & $L_1$     & $110$m            & $\overline{U}$ & $\left[ 7.0, 34^{\circ} \right]^T$ & $\underline{U}$ & $\left[ -7.0, -34^{\circ} \right]^T$            \\ 
$L_2$     & $40$m  & $T_P^1$   & $90$              & $d_S^l$        & $10m$                              & $d^2_S$        & $2.5$m                                          \\ 
$T_P^2$   & $30$   & $L_V$     & $3.5$m            & $\alpha$       & $8$                                & $T_{iter}$     & $3$                                             \\ 
$W_V$     & $1.7$m & $c_0$     & $0.6$s            & $T_a$          & $0.1$s                             & road width     & $4.5$m                                          \\ 
$c_1$     & $0.7$s & $Q_X$     & $diag\{1,1,0.0\}$ & $k_U$          & $10$                               & $M_f$          & $\left[ diag\{1,0.3\}^T, 0^{2\times 2} \right]$ \\ 
$k_X$     & $10$   & $Q_U$     & $diag\{1,0.1\}$   & CasADi         & 3.6.4                              & Python         & 3.9                                             \\ \hline
\end{tabular}
\label{sim-param}
\end{table*}

%% file: misc/table02.tex
\begin{table}[htbp]
\caption{The value of $\rho$ and $\sigma$.}
\centering
\begin{tabular}{ccccc}
\hline
Iteration & 1   & 2-13 & 14-23 & 24-33 \\ \hline
$\rho$    & 0.1 & 1.0  & 10.0  & 100.0 \\
$\sigma$  & 0.1 & 1.0  & 10.0  & 100.0 \\ \hline
\end{tabular}
\label{rho_sigma}
\end{table}

%% file: misc/table03.tex
\begin{table}[htbp]
\caption{The variation of $T_{avg}$ and $T_{total}$ with the number of vehicles.}
\centering
\begin{tabular}{ccc}
\hline
vehicle number $N$ & $T_{avg}$ & $T_{total}$ \\
\hline
3                  & 1.908     & 62.963      \\
4                  & 2.085     & 68.804      \\
5                  & 2.125     & 70.115      \\
6                  & 2.143     & 70.723      \\
7                  & 2.258     & 74.530      \\
8                  & 2.546     & 84.008      \\
9                  & 2.813     & 92.826      \\
10                 & 3.063     & 101.094     \\
\hline
\end{tabular}
\label{average-total-time}
\end{table}